\documentclass[twocolumn]{aa}
\usepackage[colorlinks,linkcolor={blue},citecolor={blue},urlcolor={red}]{hyperref}

\usepackage{graphicx}
\usepackage{txfonts}
\usepackage{xcolor}
\usepackage{color}
\usepackage{float}
\usepackage{ulem}
\usepackage{orcidlink}

\newcommand{\PSF}{\textrm{psf}}
\newcommand{\tang}{_\textrm{t}}

\begin{document}

   \title{Point-Spread Function errors for 
   weak lensing - density cross-correlations. Application to UNIONS.}


   \author{%
    Ziwen Zhang\orcidlink{0000-0002-9272-5978}
    \inst{1, 2, 3}
    \fnmsep\thanks{\email{ziwen@mail.ustc.edu.cn}}
    \and
    Martin Kilbinger\orcidlink{0000-0001-9513-7138}
    \inst{1}
    \fnmsep\thanks{\email{martin.kilbinger@cea.fr}}
    \and
    Fabian Hervas Peters\inst{1}
    \and
    Qinxun Li\orcidlink{0000-0003-3616-6486}
    \inst{4}
    \and
    Wentao Luo\orcidlink{0000-0003-1297-6142}
    \inst{2,3}
    \and
    Lucie Baumont\orcidlink{0000-0002-1518-0150}
    \inst{1}
    \and
    Jean-Charles Cuillandre\orcidlink{0000-0002-3263-8645}
    \inst{1}
    \and
    S\'ebastien Fabbro\orcidlink{0000-0003-2239-7988}
    \inst{5}
    \and
    Stephen Gwyn\orcidlink{0000-0001-8221-8406}
    \inst{5}
    \and
    Alan McConnachie\orcidlink{0000-0003-4666-6564}
    \inst{5}
    \and
    Anna Wittje\orcidlink{0000-0002-8173-3438}
    \inst{6}
    }

   \institute{%
   Université Paris-Saclay, Université Paris Cité, CEA, CNRS, AIM, 91191, Gif-sur-Yvette, France 
   \and
   CAS Key Laboratory for Research in Galaxies and Cosmology, Department of Astronomy, University of Science and Technology of China, Hefei, Anhui 230026, China
   \and
   School of Astronomy and Space Science, University of Science and Technology of China, Hefei 230026, China
   \and
   Department of Physics and Astronomy, University of Utah, Salt Lake City, Utah 84102, USA
   \and
   NRC Herzberg Astronomy \& Astrophysics, 5071 West Saanich Road, British Columbia, Canada V9E2E7
   \and
   Ruhr University Bochum, Faculty of Physics and Astronomy, Astronomical Institute (AIRUB), German Centre for Cosmological Lensing, 44780 Bochum, Germany
   }
   \date{Received; accepted}
 
  \abstract%
    {%
    }%
    {%
    Calibrating the point spread function (PSF) is a fundamental part of weak gravitational lensing analyses. Even with corrected galaxy images, imperfect calibrations can introduce biases. We propose an analytical framework for quantifying PSF-induced systematics as diagnostics for cross-correlation measurements of weak lensing with density tracers, e.g., galaxy-galaxy lensing. We show how those systematics propagate to physical parameters of the density tracers. Those diagnostics only require a shape catalogue of PSF stars and foreground galaxy positions.
    }%
    {%
    We consider the PSF-induced multiplicative bias, and introduce three second-order statistics as additive biases.
    We compute both biases for the weak-lensing derived halo mass of spectroscopic foreground galaxy samples, in particular, their effect on
    the tangential shear and fitted halo mass as a function of stellar mass. In addition, we assess their impact on the recently published black-hole - halo-mass relation for type I Active Galactic Nuclei (AGNs).
    }
    {%
    Using weak-lensing catalogues from the Ultraviolet Near Infrared Optical Northern Survey (UNIONS) and Dark Energy Survey (DES), we find the multiplicative biases in the tangential shear to be less than $0.5\%$.
    No correlations between additive bias and galaxy properties of the foreground sample are detected.
    The combined PSF systematics affect low-mass galaxies and small angular scales; halo mass estimates can be biased by up to 18$\%$ for a sample of central galaxies in the stellar mass range 9.0 $\leq$ log $M_*/\rm M_{\odot}$ < 9.5.
    }%
    {%
    The PSF-induced multiplicative bias is a subdominant contribution to current studies of weak-lensing - density cross-correlations, but might become significant for upcoming Stage-VI surveys. 
    For samples with a low tangential shear, additive PSF systematics can induce a significant bias on derived properties such as halo mass. 
    }
    

   \maketitle
%

\section{Introduction}
Light from distant galaxies on its way to the observer is affected by gravitational fields along the line of sight, distorting the light distribution of the galaxies we observe. Weak gravitational lensing refers to the typical few per cent distortions of the galaxy image due to foreground large-scale structure \citep{Jarvis2016}. Weak lensing enables us to measure the distribution of the total foreground mass, which consists of baryonic and dark matter. Therefore, it is a powerful tool for studying cosmology \citep{K15, 2017arXiv171003235M}. 

Among the main applications of weak lensing is galaxy-galaxy lensing, which is the correlation between the shapes of background galaxies and the positions of foreground galaxies. Many studies have used this method to estimate the (dark-matter) halo mass, establishing various relations between the halo mass and galaxy properties \citep{Mandelbaum2006, Leauthaud2012, Velander2014MNRAS.437.2111V, Viola2015MNRAS.452.3529V, Luo2018ApJ,Zhang2021,Zhang2022,zhang2024}.

There are a number of potential systematics that can bias weak-lensing measurements \citep{Jarvis2016}, such as the contamination from cosmic rays and satellite trails, among others, thus hindering the measurement of the galaxy brightness distribution; observed galaxy intensity profiles that may be contaminated by the flux from nearby galaxies or stars; and the observed images of galaxies that are blurred due to atmospheric refraction or turbulence and optical imperfections. The combined effect of image blurring due to the atmosphere and the optical system is known as the point spread function \cite[PSF; see for a recent focused review][]{2023FrASS..1058213L}.

The PSF can strongly smear the weak-lensing shear information in the observed galaxy shapes \citep{Jarvis2021}. In addition, the PSF is difficult to estimate because it varies with the field of view (FOV) and across individual exposures in multi-epoch observations. Therefore, a central part of weak-lensing analyses and a formidable challenge is to correct for the PSF on observed galaxy images. Nevertheless, even after PSF correction systematic errors can remain, induced by imperfections in the PSF model and the interpolation process \citep{Jarvis2016}.

Various methods exist to quantify PSF-induced systematics for weak lensing. For cosmic shear, the $\rho$-statistics \citep{2010MNRAS.404..350R, Jarvis2016} are additive biases to the shear two-point correlation function and directly propagate to cosmological parameters.
A generalisation is the so-called $\tau$-statistics \cite{DES21_Gatti} allow to separately estimate the impact of PSF leakage and PSF model errors on galaxy shapes.
PSF leakage quantifies, via the multiplicative parameter $\alpha$, how much the PSF ellipticity influences the PSF-corrected galaxy ellipticity.

For galaxy-galaxy lensing two kinds of null-test estimators have been developed. The first is the tangential shear around positions that are not correlated with the density tracers in an ideal setting. These can be
random positions \citep{2005MNRAS.361.1287M},
stars,
or points fixed to a CCD-frame coordinate system
\citep{DES21_Gatti}.
The second type of null test is the cross-component shear measured around the foreground sample, which is expected to vanish if parity is conserved.

So far, to the best of our knowledge, no PSF systematic diagnostics have been devised for lensing around foreground density tracers that directly propagate to parameters of the foreground population. Here, we developed a set of three galaxy-PSF cross-correlation functions, which we dub ``$\lambda$- statistics'', that are additive terms to the tangential shear around arbitrary density tracers, e.g.~galaxies, galaxy clusters, filaments, or voids. The $\lambda$-statistics for weak lensing of density tracers are the analogue of the $\rho$-statistics for cosmic shear.

In the study of galaxy evolution, a common approach is to relate galaxy properties to their halo mass (scaling relation), thus revealing the evolutionary paths of different galaxies in their dark-matter environment \citep{Posti2019,Zhang2021,Zhang2022,zhang2024}. We examined two scaling relations in this paper, namely the stellar-mass -- halo-mass relation and the black-hole-mass -- halo-mass relation. Various studies investigated these relations in terms of analytic models \citep{Bower2017}, abundance matching \citep{Shankar2020, Moster2010, Yang2009}, and weak lensing \citep{zhang2024,2024arXiv240210740L}. Of these, weak lensing is the most direct observational method for probing halo mass. The accuracy of scaling relations is affected by the accuracy of the halo mass estimates. In this paper, we developed a method to quantify the impact of PSF-induced systematics to the weak lensing signal and also the halo mass estimates. 

The outline of this paper is as follows.
We describe the weak-lensing catalogues and the foreground sample selections in Sect.~\ref{sec:data}.
Section \ref{sec:method} introduces the $\lambda$-statistics, and presents the methods to quantify PSF-induced multiplicative and additive biases for the weak-lensing tangential shear. This section also briefly reviews the measurement of tangential shear, and the connection of shear to halo mass.
In Sect.~\ref{sec:results}, we show the results for PSF-induced multiplicative and additive biases. In addition, we investigate the impact of PSF-induced systematics on halo mass estimation.
We discuss the implications of the $\lambda$-statistics in Sect.~\ref{sec:discussion} and summarize our results in Sect.~\ref{sec:summary}.
Throughout this paper, we assume the Planck cosmology \citep{Planck2016}: $\Omega_{\rm m} = 0.307$, $\Omega_{\rm b} = 0.048$, $\Omega_\Lambda = 0.693$ and $h = 0.678$.

\section{Observational data}
\label{sec:data}

\subsection{Weak lensing catalogues}

\subsubsection{UNIONS}
\label{sec:UNIONS}

We use the weak-lensing shear catalogue from the Ultraviolet Near-Infrared Northern Sky Survey (UNIONS). Started in 2018, UNIONS is an ongoing survey that targets 4,800 deg$^2$ in the Northern Hemisphere and covers the footprint of the Euclid survey \citep{2022A&A...662A.112E}. UNIONS combines multi-band photometric images from different telescopes. These are the Canada-France Hawai'i Telescope (CFHT) providing $u$-and $r$-band images; this part of UNIONS is called the Canada-France Imaging Survey or CFIS; the Panoramic Survey Telescope and Rapid Response System (Pan-STARRS) for the $i$-and $z$-band; and Subaru, which takes images in the $z$-band in the framework of  WISHES (Wide Imaging with Subaru HSC of the Euclid Sky), and the $g$-band Waterloo Hawai'i IfA Survey (WHIGS). Here, we only use the UNIONS $r$-band data to calculate the weak-lensing shear.
Shape measurement was performed with \textsc{ShapePipe} \citep{2022A&A...664A.141F}. A first version of the ShapePipe catalogue was presented in \cite{UNIONS_Guinot_SP}. In this paper, we use v1.3 of the catalogue, which contains $83,812,739$ galaxies 
covering $3,200$ deg$^2$ of effective sky area, which was the available data in 2022 at the time of processing. The PSF was modelled with MCCD \citep{MCCD21}, which builds a non-parametric multi-CCD model of the PSF over the focal plane. To obtain the parameters of the PSF model, stars are selected on the individual exposures. The star sample is selected on the stellar locus in the size - magnitude diagram. They are split into a training sample ($80\%$) and a validation sample ($20\%$). The PSF model is obtained by optimization using the training sample. 

\subsubsection{DES}
\label{sec:DES}
We also use the DES Y3 weak lensing catalogue in our analysis \citep{DES21_Gatti}. The catalogue contains $100,204,026$ galaxies, covering $4,139$ square degrees in the sky. Images were taken in the $g$-, $r-$, $i$-, $z-$, and $Y$-band and has a weighted source number density of $5.59$ arcmin$^{-2}$. The corresponding shape noise is $0.261$.

The DES star catalogue contains 56 million objects, which are identified as stars using the stellar locus in the magnitude range $[16.5, 22.0]$. The PSF is modelled and interpolated using the PIFF algorithm (PSFs In the Full FOV, \cite{Jarvis2021}). The parametric model for a star \textit{i} is fitted iteratively using neighbouring stars, rejecting outliers until convergence is reached. Each of the $riz$ bands is fitted separately. PIFF is intended to support fitting to the entire focal plane; however, the DES Y3 catalogue is restricted to single CCD modelling.

\subsection{Foreground galaxy catalogues}
The following subsections describe two samples of foreground galaxies, which are used for weak-lensing cross-correlation analysis in this paper.

\subsubsection{Central galaxy samples}
\label{sec:galaxy_samples}

We use galaxies from the New York University Value Added Galaxy Catalogue \citep[NYU-VAGC\footnote{http://sdss.physics.nyu.edu/vagc/},][]{Blanton-05a} of the Sloan Digital Sky Survey Data Release 7 \citep[SDSS DR7,][]{Abazajian-09}. Three selection criteria were applied to these galaxies: 1.) $r$-band Petrosian apparent magnitude $r\le 17.72$; 2.) spectroscopic redshift in the range $0.01 \le z \le 0.2$; 3.) redshift completeness $C_z > 0.7$. We only use the central galaxies in our analysis, they are defined as the most massive galaxy in a galaxy group, which are identified by the group catalogue \citep{Yang2005, Yang2007}\footnote{ https://gax.sjtu.edu.cn/data/Group.html}. NYU-VAGC provides the measurements of stellar mass ($M_*$). We cross-match our central galaxies with the MPA-JHU DR7 catalogue\footnote{https://wwwmpa.mpa-garching.mpg.de/SDSS/DR7/}, which provides galaxy star-formation rates \cite[SFR,][]{Brinchmann2004} derived from both spectroscopic and photometric data of SDSS.

In the overlapping sky area between SDSS and UNIONS, there are $126,675$ central galaxies. We divide these galaxies into three stellar-mass bins: $9.0 \leq \log M_*/\rm M_{\odot} < 9.5$, $10.0 \leq \log M_*/\rm M_{\odot} < 10.5$ and $11.0 \leq \log M_*/\rm M_{\odot} < 11.5$. We further divide each stellar-mass bin into star-forming and quenched subsamples by using the demarcation line from \cite{Bluck2016}.

\subsubsection{Active Galactic Nuclei samples}
\label{sec:AGN_samples}

The AGNs used here correspond to the type I sample described in \cite{2024arXiv240210740L}. They are collected from two catalogues. The first one is the SDSS DR16 Quasar Catalog \citep{Lyke2020}. The total number of quasars is $750,414$ with redshifts in the range $0.1 < z < 6$. Their black-hole masses were estimated by \cite{Wu2022} using the FWHM of H$\beta$, Mg II, and C IV broad emission lines from spectroscopic observations. The black-hole masses used here correspond to the H$\beta$ emission lines (see \cite{Wu2022} for more details).

The second catalogue is the SDSS DR7 AGN catalogue \citep{Liu2019}, including both quasars and Seyfert galaxies. There are $14,584$ AGNs with redshifts less than $0.35$. \cite{Liu2019} used H$\alpha$ and H$\beta$ emission lines to measure their corresponding black-hole masses. We also adopt the black-hole mass based on the H$\beta$ emission line.

We merged the two catalogues with duplicates removed, selecting AGNs in the redshift range $0.05 < z < 0.6$. 
The final catalogue of type I AGNs in the joint SDSS-UNIONS footprint contains $14,649$ objects. We divide these AGNs into low-, medium- and high-black-hole mass bins, which are $\log M_{\rm BH}/\rm M_{\odot} < 7.9$, $7.9 < \log M_{\rm BH}/\rm M_{\odot} < 8.5$ and $\log M_{\rm BH}/\rm M_{\odot} > 8.5$, respectively.

Note that the host galaxies of the selected AGNs include both central and satellite galaxies.

\section{Methods of analysis}
\label{sec:method}

This section describes the methods to estimate the impact of PSF-induced systematics on galaxy ellipticity and tangential shear $\gamma\tang$. We introduce the $\lambda$-statistics as additive PSF systematics to the tangential shear. We discuss our measurement and modelling methodology and review how physical galaxy properties such as halo mass are derived from the measured tangential shear.

\subsection{PSF error propagation}
\label{sec:psf_error_prop}

In general, the observed ellipticity $\varepsilon^{\textrm{obs}}$ of a galaxy is not an unbiased estimator of shear $\gamma$ at that position. The relation between
those quantities is, in complex notation to linear order,
\begin{equation}
    \varepsilon^{\textrm{obs}} = \varepsilon^\textrm{s} + (1+m) \gamma + c + \delta \varepsilon + \alpha \varepsilon^\PSF.
    \label{eq:epsobs}
\end{equation}
Here, $\varepsilon^\textrm{s}$ denotes the intrinsic ellipticity of the galaxy, which is assumed to have random orientation, and consequently has vanishing expectation value, $\langle \varepsilon^\textrm{s} \rangle = 0$.
$m$ and $c$ are the multiplicative and additive biases, respectively. 
The fourth term on the right-hand side, $\delta \varepsilon$, represents the residual in the PSF at the galaxy position due to errors in PSF measurement, modelling, and interpolation. The final term quantifies the leakage from the PSF ellipticity into galaxy ellipticity, which can arise, for example, from an insufficient PSF correction during galaxy shape measurement. The coefficient $\alpha$ is the leakage amplitude. 

These biases have a variety of origins \citep{Massey2013}: $m$ can arise from the misestimation of the PSF \citep{DES21_Gatti}, but this effect is typically small \citep{Massey2013}. Larger contributions come from calibration errors in the shear measurement algorithm. The potential origins of $c$ are PSF errors or the incomplete application of the charge transfer inefficiency \citep{Uitert2016, DES21_Gatti}.

In the model described by Eq.~\eqref{eq:epsobs}, we assume that $m$ and $c$ are not correlated with the PSF; we exclude PSF-induced effects from these two biases. The additive bias due to the PSF is shown in the last two terms in Eq.~\eqref{eq:epsobs}. We will show below how the PSF uncertainty induces an additional multiplicative bias that is not included in the above equation. Since we are primarily focusing on PSF errors, we assume that the shear measurements have been accurately calibrated for $m$ and $c$, and set $m = c = 0$.

Assuming ellipticity to be measured via second moments of the light distribution, the PSF residual has been derived via Gaussian error propagation in \cite{Henriksson2008} as
\begin{equation}
    \delta \varepsilon
        = \left( \varepsilon^{\rm obs} - \varepsilon^\PSF \right)
        \frac{\delta T^\PSF}{T}
        - \frac{T^\PSF} T \delta \varepsilon^\PSF,
    \label{eq:delta_eps_original}
\end{equation}
where $T$ is the galaxy's intrinsic (PSF de-convolved) size, and $T^\PSF$ is the PSF size. The PSF residual is induced by the error of the PSF model, which is denoted as the difference between the measured and modelled PSF in their size, $\delta T^\PSF$, and ellipticity, $\delta \varepsilon^\PSF$. Unfortunately, we can not directly measure $\delta T^\PSF$ and $\delta \varepsilon^\PSF$ at galaxy positions. However, the PSF can be estimated using stars, and PSF residuals can be obtained at star positions. Therefore, we write eq.~\eqref{eq:delta_eps_original} as
%
\begin{equation}
    \delta \varepsilon
        = \left( \varepsilon^\textrm{obs} - \varepsilon^\PSF \right)
        \frac{T^\PSF} T \frac{\delta T^\PSF}{T^\PSF}
        - \frac{T^\PSF} T \delta \varepsilon^\PSF,
    \label{eq:delta_eps}
\end{equation}
and measure $T^\PSF / T$ at galaxy positions, and $\delta T^\PSF /T^\PSF$ at star positions.

\subsection{PSF-induced systematics for tangential shear}
\label{sec:lambda_statistics}

A sample $S$ of objects that trace matter in the large-scale structure induces a tangential shear $\gamma\tang^S$ on background galaxies. This tangential shear can be written as second-order correlation between background shear and foreground number density $n$. Therefore, an average tangential shear caused by a foreground sample $S$ is
\begin{equation}
    \gamma\tang^S =
        \left\langle
            \gamma\tang \, n 
        \right\rangle .
    \label{eq:gamma}
\end{equation}
An estimator of this tangential shear is
\begin{equation}
    \hat \gamma\tang \equiv \gamma\tang^{\rm obs} =
        \left\langle
            \varepsilon\tang^\textrm{obs} n 
        \right\rangle .
    \label{eq:gamma_hat}
\end{equation}
This estimator is the average of the observed tangential ellipticity of background galaxies $\varepsilon\tang^\textrm{obs}$ around foreground positions.

We can additionally define the cross-component of shear, $\gamma_\times$, rotated by $45$ degrees with respect to $\gamma\tang$. These two components can be combined to form the complex shear $\gamma = \gamma_\textrm{t} + \textrm{i} \gamma_\times$. The tangential shear $\gamma_{\textrm{t}}$ is identified as $E$-mode, induced via gravitational lensing, whereas the cross-component $\gamma_\times$ indicates the parity-odd $B$-mode.

We denote $\delta \gamma\tang$ with
\begin{equation}
    \delta \gamma\tang = \gamma^{\rm obs}\tang - \gamma^{S}\tang,
    \label{eq:dg_gamma}
\end{equation}
the difference between the estimated and true shear.
Inserting Eq.~\eqref{eq:epsobs} into Eq.~\eqref{eq:gamma_hat}, we find for the residual tangential shear component to be
\begin{align}
    \delta \gamma\tang
        = \left\langle \varepsilon^\textrm{s}\tang n \right\rangle +
        & \left\langle
            \frac{T^\PSF} T \frac{\delta T^\PSF}{T^\PSF} 
          \gamma\tang n \right\rangle
        - \left\langle
            \frac{T^\PSF} T
            \frac{\delta T^\PSF} {T^\PSF} \varepsilon\tang^\PSF n
          \right\rangle
        \nonumber \\
        & - \left\langle
            \frac{T^\PSF} T
            \delta \varepsilon\tang^\PSF n
            \right\rangle
        + \left\langle \alpha
            \varepsilon\tang^\PSF n
            \right\rangle.
\end{align}
The correlators can be further expanded under some assumptions, as follows. First, we can safely assume that the intrinsic galaxy ellipticity is uncorrelated with foreground number density.
Next, we assume that the PSF model uncertainties are not correlated to the shear. 
Third, we separate out the prefactors $T^\PSF / T$ and $\alpha$, following \cite{2010MNRAS.404..350R}. This yields
\begin{align}
    \delta \gamma\tang
        = & 
        \left\langle 
                \frac{T^\PSF} T
            \right\rangle
            \left\langle
                \frac{\delta T^\PSF}{T^\PSF} 
          \right\rangle \gamma^{S}\tang
        - \left\langle
            \frac{T^\PSF} T
          \right\rangle
            \left\langle 
                \frac{\delta T^\PSF} {T^\PSF} \varepsilon\tang^\PSF n
           \right\rangle 
        \nonumber \\
        & - \left\langle
                \frac{T^\PSF} T
            \right\rangle
            \left\langle
                \delta \varepsilon\tang^\PSF n
            \right\rangle
        + \alpha
          \left\langle 
            \varepsilon\tang^\PSF n
          \right\rangle.
\end{align}

We define three new cross-correlation functions in analogy to the $\rho$-statistics introduced for cosmic shear in \cite{Jarvis2016}. These functions are
\begin{align}
    \lambda_1
        & = \left\langle
                \varepsilon\tang^\PSF \, n
            \right\rangle ;
    \nonumber \\
    \lambda_2
    & = \left\langle
            \frac{\delta T^\PSF} {T^\PSF} \varepsilon\tang^\PSF \, n
        \right\rangle;
    \nonumber \\
    \lambda_3
    & = \left\langle
            \delta \varepsilon\tang^\PSF \, n
        \right\rangle .
    \label{eq:lambda_123}
\end{align}
With that, we write the shear difference as
\begin{equation}
   \delta \gamma\tang =
        \left\langle
            \frac{T^\PSF} T
        \right\rangle
        \left\langle
            \frac{\delta T^\PSF}{T^\PSF} 
        \right\rangle \gamma^{S}\tang
     + \alpha \lambda_1
     - \left\langle
            \frac{T^\PSF} T
       \right\rangle
        \left(
            \lambda_2 + \lambda_3
        \right). 
    \label{eq:dg_lambda_theory}
\end{equation}
The first term on the right-hand side, the prefactor of $\gamma_{\rm t}^{S}$, is a PSF-induced multiplicative bias. The remaining three terms are PSF-induced additive biases, expressed as the $\lambda$-statistics, i.e., the correlations between the PSF and foreground positions. We use the TreeCorr\footnote{https://pypi.org/project/TreeCorr/} package \citep{Jarvis2015} to calculate the $\lambda$ statistics. Their error bars are estimated using the jackknife method. 

In the case of cosmic shear, the multiplicative term is usually ignored, and only the additive terms to the shear two-point correlation function $\xi_+$ are kept. In the following, we will consider both contributions using two different approaches. The first approach is indicated in Eq.~\eqref{eq:dg_lambda_theory}, which assumes the knowledge of $\gamma^{S}\tang$. However, $\gamma^{S}\tang$ is not a direct observable. Therefore, we adopt a theoretical model to derive $\gamma^{S}\tang$ given a sample of foreground tracers $S$ (see Sect.~\ref{sec:M1_halo_mass} for details). For the second approach, we calculate $\delta\gamma\tang$ from the observation by combing Eqs.~\eqref{eq:dg_gamma} and \eqref{eq:dg_lambda_theory}. Consequently, $\delta\gamma\tang$ can be written as
\begin{equation}
\label{eq:dg_lambda_obs}
   \delta\gamma\tang = \gamma\tang^{\rm obs} - \frac
   {
        \gamma\tang^{\rm obs}
        - \alpha \lambda_1
        +\left\langle
            \frac{T^\PSF} T
       \right\rangle
        \left(
            \lambda_2 + \lambda_3
        \right) \
   }
   {
        1 
        + \left\langle
            \frac{T^\PSF} T
        \right\rangle
        \left\langle
            \frac{\delta T^\PSF}{T^\PSF} 
        \right\rangle
    },
\end{equation}
where $\gamma\tang^{\rm obs}$ is measured by using the shear catalogue around foreground tracers $S$ (see Section \ref{sec:M2_gt_obs} for the calculation of $\gamma_{\rm t}^{\rm obs}$). 
In the limit where PSF-induced multiplicative bias vanishes, this expression reduces to Eq.~\eqref{eq:dg_lambda_theory}.

We refer to the two expressions Eqs.~\eqref{eq:dg_lambda_theory} and \eqref{eq:dg_lambda_obs} as the theory-based and observation-based approach, respectively. We denote the corresponding residual tangential shear as $\delta\gamma_{\rm t}^{\rm theory}$ and $\delta\gamma_{\rm t}^{\rm obs}$, respectively.

\subsection{Theory-based approach and its impact on constraining halo mass}
\label{sec:M1_halo_mass}

The key ingredient of the theory-based approach, Eq.~\eqref{eq:dg_lambda_theory}, in calculating the residual tangential shear $\delta\gamma_{\rm t}^{\rm theory}$ is the derivation of the model prediction $\gamma^{S}\tang$. We use an analytical method based on three assumptions. All galaxies from the foreground sample described in Sect.~\ref{sec:galaxy_samples} are central galaxies. We assume that these galaxies follow the stellar mass - halo mass relation (SHMR) from \cite{Kravtsov2018}. We calculate the average stellar mass of the galaxy sample and interpolate the SHMR to infer the average halo mass of the sample. We also assume that the halo is described by the Navarro–Frenk–White \cite[NFW]{Navarro1997} density profile and define the halo mass as the total mass within a spherical region of radius $r_{200 {\rm m}}$, inside which the mean mass density is equal to $200$ times the mean matter density of the Universe. We further model the halo mass-halo concentration relation with the fit from \cite{Bhattacharya2013ApJ}. With these assumptions, we can use the analytical equations from \cite{Yang2006} to derive the excess surface density, $\Delta \Sigma\tang$. 

The relation between $\Delta \Sigma\tang$ and $\gamma^{S}\tang$ is given by the following equations
\begin{align}
    \Delta \Sigma \tang & = \Sigma_{\rm crit}\ \gamma^{S}\tang;
    \nonumber \\
    \Sigma_{\rm crit}(z_{\rm l}, z_{\rm s}) & = \frac{c^{2}}{4\pi G}\frac{D_{\rm s}}{D_{\rm l}D_{\rm ls}},
\end{align}
where $c$ is the speed of light, $G$ is the gravitational constant, $D_{\rm l}$, $D_{\rm s}$ and $D_{\rm ls}$ are the angular diameter distances to the lens, to the source and between the lens and source, respectively. We adopt the mean redshift of the foreground galaxy sample to calculate $z_{\rm l}$ and the corresponding $D_{\rm l}$. Instead of using a single source redshift value for $z_{\rm s}$, we use the redshift distribution $n(z)$ of the UNIONS catalogue to derive $\Sigma_{\rm crit}$ via
\begin{equation}
    \Sigma^{-1}_{\rm crit}(z_{\rm l}) = 
    \int_{z_{\rm l}}^{z_{\rm lim}}
    \Sigma_{\rm crit}^{-1}
        (z_{\rm l}, z_{\rm s}) \,
    n(z_{\rm s}) \, {\rm d} z_{\rm s},
    \label{eq:sigma_cr}
\end{equation}
where $z_{\rm lim}$ represents the maximum redshift value in the redshift distribution of the UNIONS catalogue.

Taken together, we can derive $\gamma^{S}\tang$ and hence $\delta\gamma_{\rm t}^{\rm theory}$. The above approaches were referred to as the Yang-halo model. Finally, we combine $\delta\gamma_{\rm t}^{\rm theory}$ and $\gamma^{S}\tang$ to infer $\hat\gamma^{\rm obs}_{\rm t}$, which represents the estimator of the observed tangential shear after the true tangential shear has been contaminated by PSF-induced systematics.

To investigate the influence of the $\delta\gamma_{\rm t}^{\rm theory}$ on the estimated halo mass, we constrain the halo mass by fitting $\hat\gamma^{\rm obs}_{\rm t}$ using a modified version of the Yang-halo model: instead of using the halo mass - halo concentration relation from \cite{Bhattacharya2013ApJ}, we set these two parameters as the free parameters in the model. The priors of these parameters are chosen to be flat, with the logarithm of the halo mass (log$M_{\rm h}$/M$_{\odot}$) in the range of [10.0, 16.0] and halo concentration in the range of [1.0, 16.0]. To constrain these parameters, we use $emcee$\footnote{https://emcee.readthedocs.io/en/stable/} \citep{Foreman-Mackey2013PASP} to run a Markov Chain Monte Carlo (MCMC). 
The likelihood function is set to be
\begin{equation}
\label{eq:likeLihood}
    \ln \mathcal L  = -{1\over 2} \left( \hat\gamma^{\rm obs}\tang - \gamma^{\rm model}\tang \right)^\textrm{T} C^{-1} \left( \hat\gamma^{\rm obs}\tang - \gamma^{\rm model}\tang \right),
\end{equation}
where $\gamma^{\rm model}\tang$ is the model prediction and $C^{-1}$ is the inverse of the covariance matrix.

In the following, the constrained halo masses are the medium values of the posteriors, and the error bars correspond to $16\%$ and $84\%$ of the posterior distributions.

\subsection{Observation-based approach and its impact on constraining halo mass}
\label{sec:M2_gt_obs}

To apply the observation-based approach to study the PSF-induced systematics, we need to obtain the observed tangential shear $\gamma^{\rm obs}_{\rm t}$ for a galaxy sample, as shown in Eq.~\eqref{eq:dg_lambda_obs}. We use the UNIONS shear catalogue to calculate $\gamma^{\rm obs}_{\rm t}$ based on galaxy samples with different properties. The ellipticities of galaxies in UNIONS are calibrated, therefore, the following equation is accurate enough to perform the calculation
\begin{equation}\label{eq:gamma_t}
    \gamma^{\rm obs}\tang
        = \frac{\sum_i^N
            w_i \, \varepsilon_{i, \rm t}}
            {\sum_i^N w_i},
\end{equation}
where $w_i$ and $\varepsilon_{i, \textrm{t}}$ are the weight and the ellipticity along the tangential direction around the lens of the source galaxy with index $i$, respectively. The sum is carried out over all suitable source galaxies. We calculate $\varepsilon_\textrm{t}$ as
\begin{equation}
    \varepsilon_\textrm{t} =
      -\varepsilon_1
      \cos 2\theta
      -
      \varepsilon_2
      \sin 2\theta
      ,
\end{equation}
where $\varepsilon_\textrm{1}$ and $\varepsilon_\textrm{2}$ are two ellipticity components of the source galaxy, and $\theta$ is the angle between the line connecting the lens and the source and the direction of increasing right ascension. 

The source galaxies in UNIONS do not have photometric redshifts, and we use all sources in our calculations. Some source galaxies are thus in front of the ``foreground'' galaxies and will dilute the observed tangential shear. Eq.~\eqref{eq:sigma_cr} models this dilution correctly.

Based on the UNIONS shear catalogue and a given foreground galaxy sample, we apply TreeCorr to calculate its $\gamma^{\rm obs}\tang$, the error bars are estimated using the jackknife method.

We further quantify the impact of the observation-based approach on halo mass estimates. More specifically, we investigate the impact of PSF-induced systematics on black-hole-mass - halo-mass relation. \cite{2024arXiv240210740L} recently investigated this relation for type I AGNs using the UNIONS shear catalogue; we adopt the method in \cite{2024arXiv240210740L} to calculate $\gamma^{\rm obs}\tang$ and therefore $\Delta \Sigma_\textrm{t}$. Then, we derive $\delta\gamma_{\rm t}^{\rm obs}$ by Eq.~\eqref{eq:dg_lambda_obs}. The $\hat\gamma^S\tang$ was obtained by combining $\gamma^{\rm obs}\tang$ and $\delta\gamma_{\rm t}^{\rm obs}$, which represents the estimator of the true tangential shear without PSF-induced systematics.

We use the halo model introduced in \cite{Guzik2002} and applied in \cite{2024arXiv240210740L} to fit the excess surface mass density $\Delta \Sigma_\textrm{t}$ and $\Sigma_{\rm crit}\hat\gamma^S\tang$ of the AGN sample, respectively. Our model contains contributions of central host halo, $\Delta \Sigma_\textrm{cen}$, and satellite host halo, $\Delta \Sigma_\textrm{sat}$. It further includes a baryonic term from the galaxy, $\Delta \Sigma_{\rm b}$, and a two-halo term, $\Delta \Sigma_{\rm 2h}$. The overall model is
\begin{equation}\label{eq:AGN_halo_model}
    \Delta \Sigma_\textrm{t} = (1-f_{\rm sat})\ \Delta \Sigma_{\rm cen} + f_{\rm sat}\ \Delta \Sigma_{\rm sat} + \Delta \Sigma_{\rm b} + \Delta \Sigma_{\rm 2h},
\end{equation}
where $f_{\rm sat}$ describes the fraction of satellite galaxies in the sample.

We treat the baryonic contribution of the galaxy as a point mass and consider the stellar mass as a free parameter in the model. The two-halo term is calculated by combining the halo bias model from \cite{Tinker2010} and the linear matter-matter correlation function from \textsc{COLOSSUS} \citep{Diemer2018ApJS}. The model contains a total of three free parameters, which are the halo mass ($M_\textrm{h}$), the stellar mass ($M_\ast$), and the satellite fraction ($f_{\rm sat}$), respectively.
We constrain these parameters by running MCMC with the $emcee$ package. The likelihood function is the same as Eq.~\eqref{eq:likeLihood}. We refer to the above model as the AGN-halo model. See 
\cite{2024arXiv240210740L} for more details on calculating and modelling the weak lensing signals of the AGN samples.

\subsection{Halo mass change and tangential shear residuals}
\label{sec:Mh_residual}

The average halo mass of a galaxy sample derived from the tangential shear $\hat\gamma^{\rm obs}_{\rm t}$ as described in Sect.~\ref{sec:M1_halo_mass} might be biased by PSF-induced systematics. To visualize this potential change in halo mass, we define a halo mass $M_\textrm{h}^P$ modified by a percentage $P$ as
\begin{equation}
    M_\textrm{h}^P = M_\textrm{h} \cdot (1 + P).
\end{equation}
where $M_\textrm{h}$ corresponds to the halo mass derived from the SHMR. In this way, we can evaluate the percentage change in halo mass corresponding to PSF-induced residual tangential shear $\delta \gamma\tang(\theta)$ at different angular scales $\theta$. As before, we use the Yang-halo model to generate the tangential shears corresponding to different $M_\textrm{h}^P$.

\section{Results}
\label{sec:results}

In this section, we present our results on the impact of PSF-induced multiplicative and additive biases on galaxy-galaxy lensing.

\subsection{PSF-induced multiplicative bias}
\label{sec:mul_bias}

As introduced in Eq.~\eqref{eq:dg_lambda_theory}, the multiplicative bias associated with the PSF manifests itself as a prefactor. 
Table \ref{tab:prefactor} shows the average prefactors for both the UNIONS and DES catalogues. The PSF leakage $\alpha$ is typically computed via ratios of galaxy- and PSF auto- and cross-correlations. We quote its values from \cite{UNIONS_Guinot_SP} and \cite{DES21_Gatti} for the UNIONS and DES catalogues, respectively.

\begin{table}[H]
    \centering
    \caption{Prefactors and PSF leakage $\alpha$ for different catalogues.}
    \renewcommand{\arraystretch}{1.2}
    \setlength{\tabcolsep}{6.5mm}{
    \begin{tabular}{c c c}
    \hline
      Parameter  & UNIONS  & DES \\
    \hline
    $\left\langle\frac{T^{\rm psf}}{T}\right\rangle$                                                                & $1.3189$ & $1.014$ \\\
    $\left\langle\frac{\delta T^{\rm psf}}{T^{\rm psf}}\right\rangle$                                               & $-0.0032$ & $0.0003$ \\\
    $\left\langle\frac{T^{\rm psf}}{T}\right\rangle\left\langle\frac{\delta T^{\rm psf}}{T^{\rm psf}}\right\rangle$ & $-0.0042$ & $0.0003$ \\\
    $\alpha$ & $0.033$ & $0.001$ \\
    \hline
    \end{tabular}
    }
    \renewcommand{\arraystretch}{1}
    \label{tab:prefactor}
\end{table}

The PSF-induced multiplicative biases in both catalogues are lower than 1$\%$, indicating that their PSFs are well calibrated. 
This bias is thus smaller than the typical residual multiplicative bias of current weak-lensing surveys, which is of order $1$--$2\%$ \cite[e.g.][]{2021A&A...645A.105G,2022MNRAS.509.3371M}. We conclude that the PSF uncertainty is not a major contributor to the multiplicative bias.

\subsection{PSF-induced additive bias}
\label{sec:additive_bias}

To examine the overall intrinsic bias levels of the UNIONS and DES catalogues, we randomly generate $2,000,000$ positions in the survey footprints. We use these random samples as the position catalogue $n$ in Eq. \ref{eq:lambda_123}. This gives us an indication of a correlation between the PSF and the footprint mask.

\begin{figure}[H]
    \centering
    \includegraphics[scale=0.16]{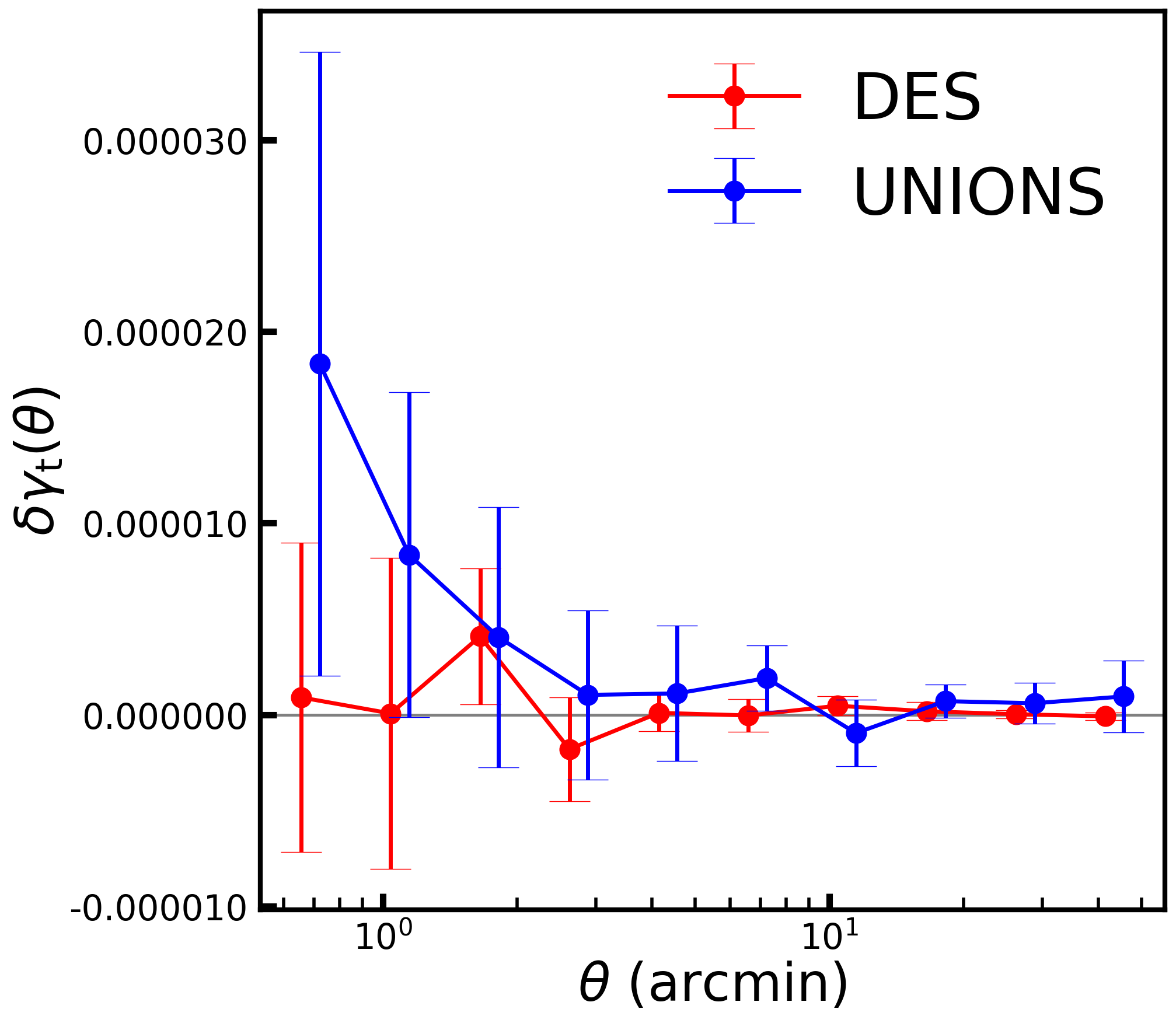}
    \caption{Comparison of PSF-induced additive bias for the DES and UNIONS catalogues based on random positions. The bias for DES (UNIONS) is shown in red (blue). The error bars correspond to the 1$\sigma$ uncertainty using the jackknife method.}
    \label{fig:lam_random}
\end{figure}

We show the residual tangential shear (additive terms of Eq. \ref{eq:dg_lambda_theory}) in Fig.~\ref{fig:lam_random}.
The PSF-induced additive biases for UNIONS and DES are broadly consistent with zero within the margin of error. 
The overall amplitude of PSF-induced additive biases is below $2 \cdot 10^{-5}$ for both surveys, with DES displaying a lower level of systematic contributions. The consistency with zero on most angular scales indicates an accurate PSF correction for both surveys. However, in regimes where the tangential shear induced by a density tracer is below a few times of $10^{-6}$, the gravitational lensing shear might be significantly affected by PSF errors.

Previous weak-lensing-based studies have investigated the relation between galaxy stellar mass and halo mass and have found a positive correlation between these two quantities. The dispersion in this relation is however large. In addition, at fixed stellar mass, quenched galaxies tend to reside in more massive halos compared to star-forming galaxies \citep{Mandelbaum2016, Bilicki2021, Zhang2021, zhang2024}. Here, we investigate whether PSF-induced systematics can bias the relation between galaxy properties and halo mass. 

Galaxy samples have selection functions that depend on their properties, e.g. size, magnitude or star-formation rate. The galaxy selection can vary with observing conditions such as seeing, image quality, local star and galaxy density. Since these factors also depend on the PSF, the galaxy-galaxy lensing PSF systematics discussed in this work can vary between different foreground samples.

We calculate the dependence of PSF-induced additive bias on the foreground galaxy samples with different stellar masses as well as SFR (see Sect.~\ref{sec:galaxy_samples} for the sample selections). 
We focus on the UNIONS catalogue and its large overlap with SDSS/eBOSS.

\begin{figure*}
    \centering
    \includegraphics[scale=0.12]{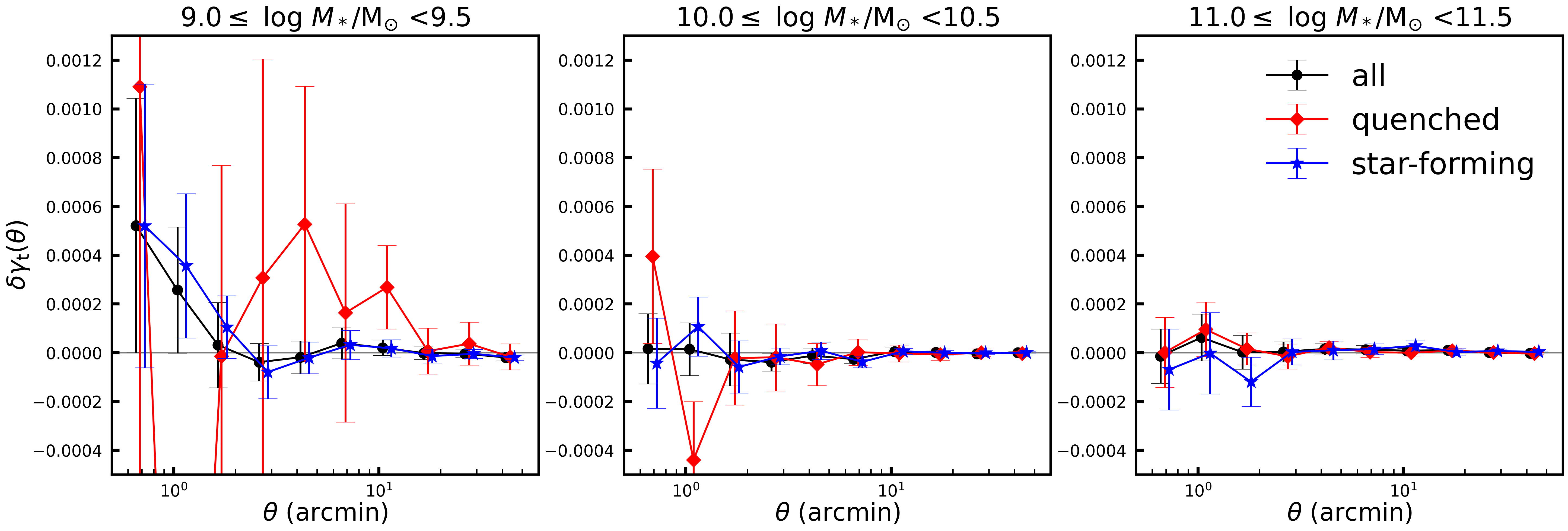}
    \caption{The PSF-induced additive bias for the central galaxy samples with different stellar mass and SFR. The three panels correspond to the three stellar-mass bins. In each panel, black circles show all galaxies in the stellar-mass bin, while the red diamonds (blue stars) correspond to the quenched (star-forming) subsample. Weak-lensing measurements use the UNIONS data.}
    \label{fig:lam_parameter}
\end{figure*}

The PSF-induced additive biases based on different stellar-mass samples are shown in Fig.~\ref{fig:lam_parameter}. In each stellar-mass bin, PSF-induced bias is consistent with zero on most angular scales; on small scales (below $1.3$ arcmin) there is a large scatter. This indicates that PSF-induced systematics do not have a significant effect on the galaxy-galaxy lensing of these samples.

Also shown are PSF-induced additive biases for subsamples with different SFR.  Overall, the biases of the quenched and star-forming subsamples are consistent within the error bars. 
We do not find a significant dependence of PSF-induced additive bias on stellar mass and SFR. 

\subsection{The overall PSF-induced systematics on weak-lensing measurements}

We now turn towards the overall effect of PSF-induced biases on weak lensing measurements and halo mass estimates. We used stellar-mass samples described in section \ref{sec:additive_bias} as density tracers and the UNIONS star and galaxy catalogues as shape catalogues.

\begin{figure*}
    \centering
    \includegraphics[scale=0.12]{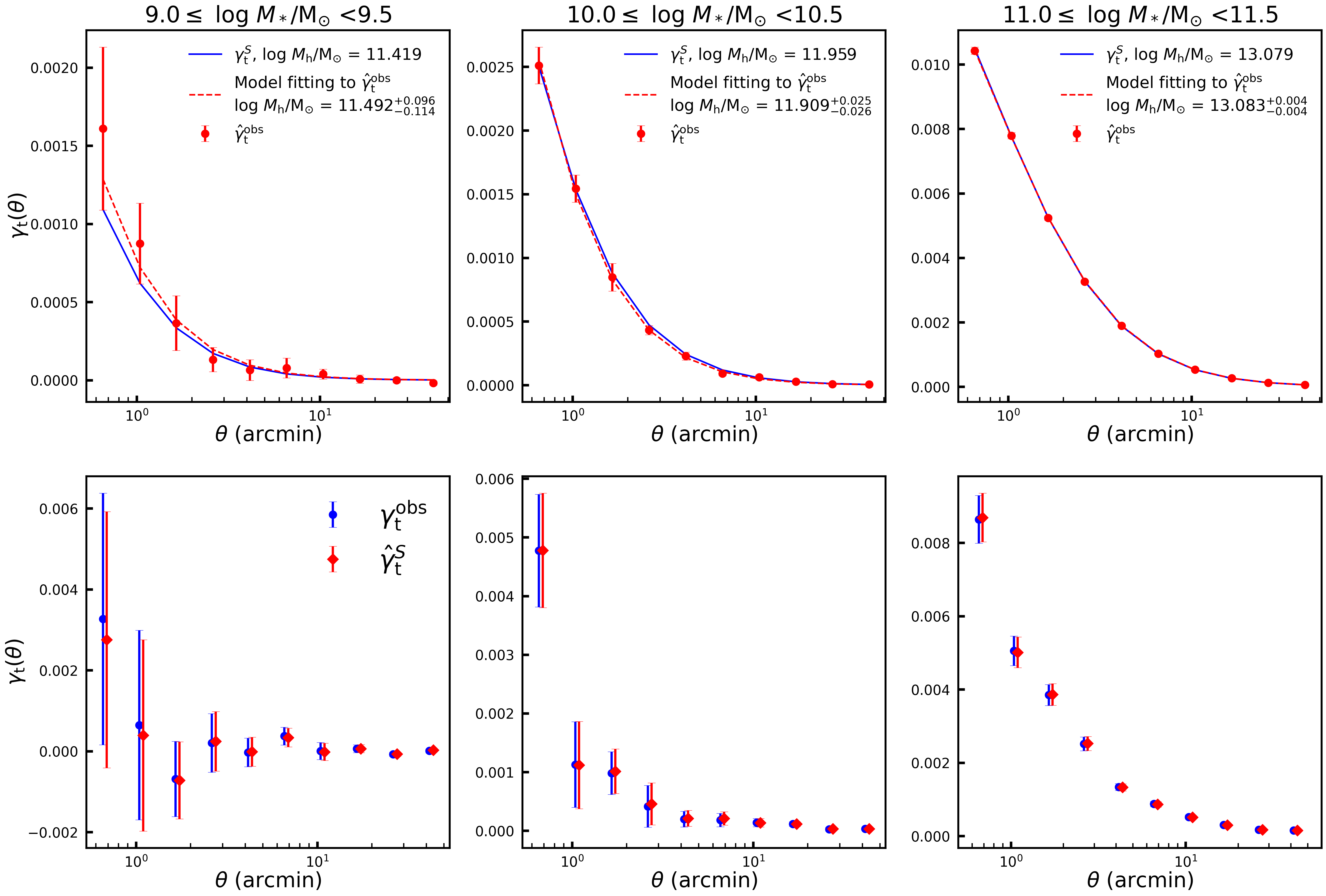}
    \caption{The effect of PSF-induced systematics in UNIONS weak lensing by the theory-based and observation-based approaches. Different columns correspond to the results of galaxy samples in different stellar mass bins. In each upper panel, the blue solid line is the theoretical $\gamma\tang^S$, while the red circles with error bars and the red dashed line are the 'observed' tangential shear ($\hat\gamma_{\rm t}^{\rm obs}$) and model fitting, respectively. Their corresponding halo masses are also labelled in the panel. In each lower panel, the blue and red symbols with error bars correspond to the observed tangential shear ($\gamma\tang^{\rm obs}$) calculated from the UNIONS shear catalogue and the estimator of the true tangential shear ($\hat\gamma\tang^S$), respectively.}
    \label{fig:dg}
\end{figure*}

We first applied the theory-based approach introduced in Sect.~\ref{sec:M1_halo_mass} to predict the theoretical tangential shear $\gamma\tang^S$ of the foreground galaxy sample. With that we calculated the residual tangential shear $\delta\gamma_{\rm t}^{\rm theory}$ using Eq.~\eqref{eq:dg_lambda_theory}. The observed shear $\hat\gamma_{\rm t}^{\rm obs}$ was computed by combining $\gamma\tang^S$ and $\delta\gamma_{\rm t}^{\rm theory}$, which now includes PSF-induced systematics. We then fitted to the $\hat\gamma_{\rm t}^{\rm obs}$ by the model described in Sect.~\ref{sec:M1_halo_mass}.
By comparing the best-fit halo mass with the corresponding halo mass of $\gamma\tang^S$, we can quantify the difference in halo mass due to PSF-induced systematics. The results are shown in the upper panels of Fig.~\ref{fig:dg}.

The tangential shear is most affected by the PSF in the lowest stellar mass bin, and mainly on small scales. The corresponding halo mass is biased by up to 18$\%$. For higher stellar masses, the measured halo masses deviate by 11$\%$ and 1$\%$, respectively. 

Next, we applied the observation-based approach. We calculated the observed tangential shear $\gamma^{\rm obs}_{\rm t}$ for the above galaxy samples, computed $\delta\gamma_{\rm t}^{\rm obs}$ using Eq.~\eqref{eq:dg_lambda_obs} and $\hat \gamma^{S}_{\rm t}$ from Eq.~\eqref{eq:dg_gamma}. The comparisons are shown in the lower panels of Fig.~\ref{fig:dg}. Similar to the theory-based approach, PSF-induced systematics mainly affect the results in the low stellar mass range at a level which is smaller than the statistical errors.

\begin{figure*}
    \centering
    \includegraphics[scale=0.245]{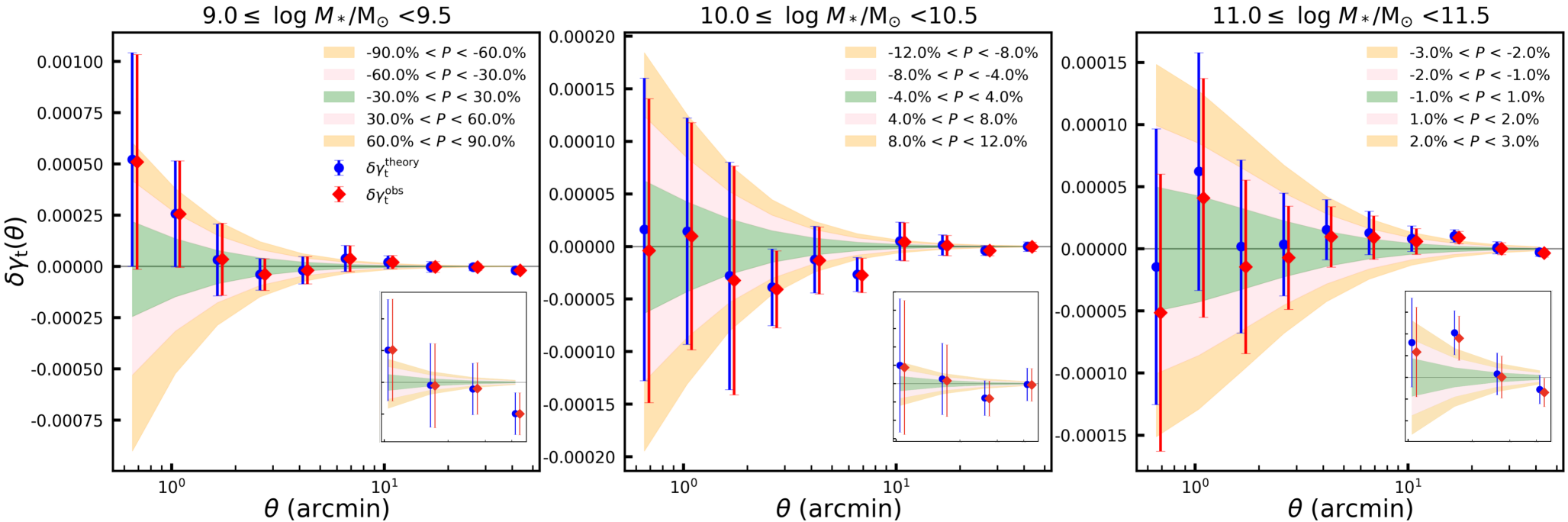}
    \caption{Comparison of $\delta\gamma_{\rm t}^{\rm theory}$ and $\delta\gamma_{\rm t}^{\rm obs}$ and the halo mass-induced residual tangential shears from UNIONS. Different panels correspond to the results in different stellar mass bins. In each panel, the blue and red symbols with error bars correspond to the $\delta\gamma_{\rm t}$ from the theory-based and observation-based approaches, respectively. Shaded regions of different colors correspond to the different degrees of the halo mass-induced shear residuals. The subplot in each panel is the zoomed-in view of the large-scale signal.}
    \label{fig:compare}
\end{figure*}

We compared PSF-induced systematics from the two approaches and also the residual tangential shear due to halo mass change (For calculating the halo mass-induced residual tangential shear, see Sect. \ref{sec:Mh_residual}) in Fig.~\ref{fig:compare}. The theory- and observation-based approaches yield similar results. As discussed in Sect.~\ref{sec:additive_bias}, PSF-induced multiplicative bias is very small so that the residual tangential shear $\delta\gamma_{\rm t}$ is dominated by the $\lambda$ statistics (PSF-induced additive bias).
We further compared the halo mass-induced residual tangential shear with the PSF-induced residual tangential shear. On individual angular scales, PSF-induced systematics can impact the tangential shear corresponding to a change in halo mass significantly exceeding the average bias of the halo mass. Such residuals can be removed using the methods described in this work.

Our analysis suggests that the weak-lensing measurements for the low-mass galaxies are facing potential challenges from PSF-induced systematics. In addition, the galaxy number density of these galaxies is low. The current survey depths are insufficient to detect more low-mass central galaxies. Together, these drawbacks make the accuracy of the weak-lensing measurements at the low-mass end always low. The error bars in the lensing signals are more significant than the PSF-induced systematics we detect here, making PSF-induced systematics less important. However, our work may inspire future studies using deeper data, for which correcting for PSF-induced systematics may become important.

\subsection{Impact of PSF-induced systematics on black-hole-mass - halo-mass relation}
\label{sec:Mh_AGN}

Recently, \cite{2024arXiv240210740L} studied the black-hole-mass - halo-mass relation based on SDSS AGNs as foreground samples and UNIONS as background shear catalogue. We now quantify the impact of PSF-induced systematics on the halo mass estimates for the type I AGN samples.

\begin{figure*}
    \centering
    \includegraphics[scale=0.15]{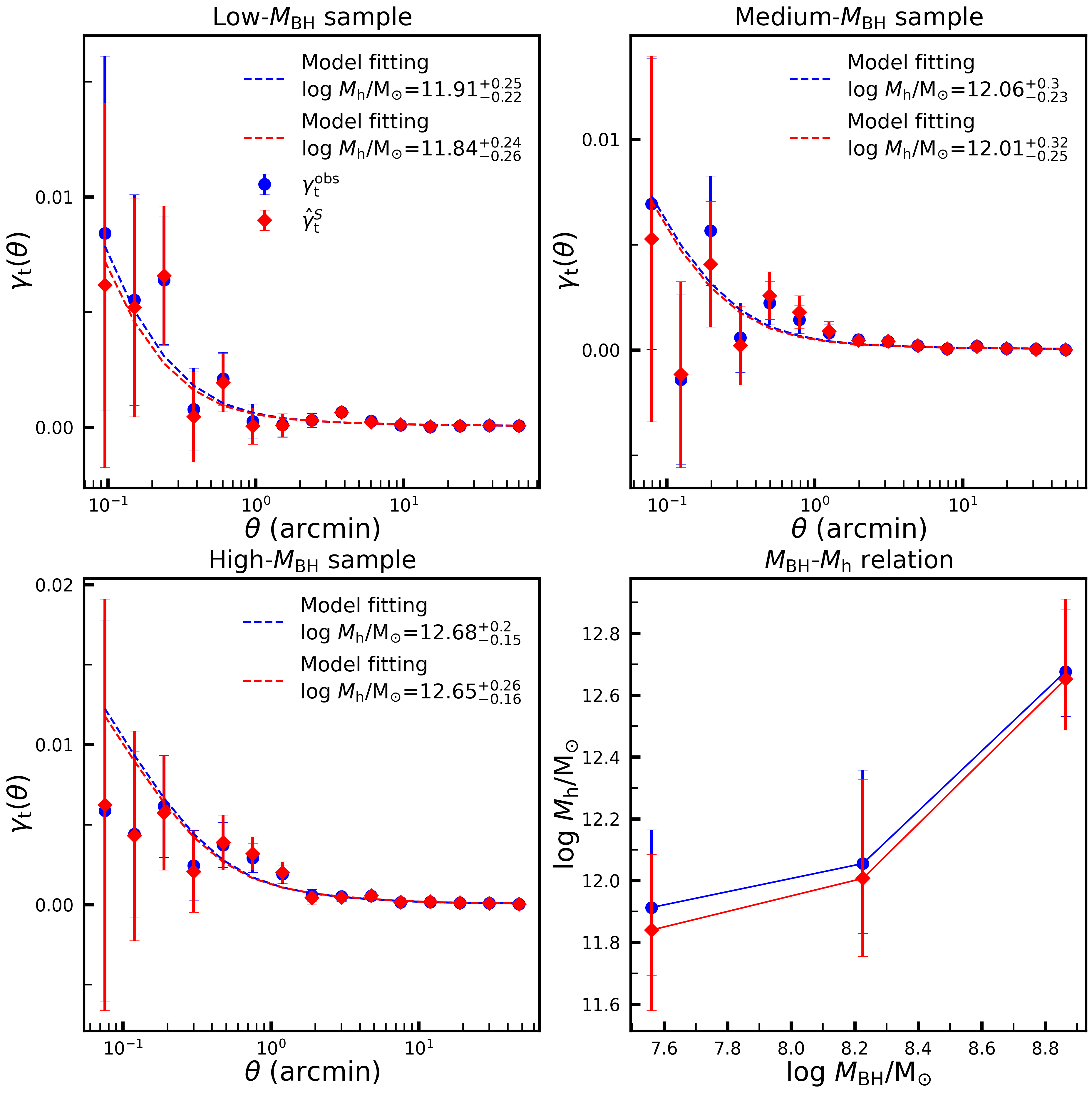}
    \caption{The effect of PSF-induced systematics on the black-hole-mass - halo-mass relation as measured with UNIONS. The upper-left, upper-right, and lower-left panels correspond to the low, medium, and high-black-hole mass AGN samples, respectively. In each panel, blue (red) symbols show the raw measured $\gamma_{\rm t}^{\rm obs}$ (PSF-induced systematics corrected $\hat\gamma_{\rm t}^{S}$) tangential shear; lines correspond to the best-fit, with the corresponding best-fit halo mass indicated in the panel. The lower-right panel shows the black-hole-mass - halo-mass relation using the raw measured (PSF-induced systematics corrected) halo masses in blue (red).}
    \label{fig:AGN}
\end{figure*}

We begin by constructing the foreground samples as in \cite{2024arXiv240210740L} (see Section \ref{sec:AGN_samples} for the sample selections). We then calculate the residual tangential shears for these samples using the observation-based approach (Eq.~\eqref{eq:dg_lambda_obs}). The results with and without the inclusion of PSF-induced systematics are shown in Fig.~\ref{fig:AGN}. 

The weak-lensing tangential shears in the three black-hole mass bins are weakly affected by PSF-induced systematics on scales larger than $2$ arcmin. The largest differences appear in the low-black-hole-mass sample.
We use the AGN-halo model (see Section \ref{sec:M2_gt_obs}) to measure the halo masses for the three AGN samples, their black-hole-mass - halo-mass relations are shown in the lower-right panel of Figure \ref{fig:AGN}.
Ignoring PSF-induced biases leads to a slight but systematic underestimation of the halo masses in the low and medium-black-hole mass samples, with the degree of deviation of their halo masses being $15\%$ and $11\%$, respectively. These halo mass deviations are however smaller than the statistical errors. 

\section{Discussion}
\label{sec:discussion}

\subsection{Motivation for the $\lambda$-statistics}

In analyses of weak-lensing correlations with density tracers, e.g. weak lensing of clusters or galaxy-galaxy lensing, null tests and diagnostics of additive biases
are used routinely. These are, for example, the mean tangential shear around non-tracers such as random points, stars, or coordinates relative to the CCDs, and the cross-component of shear around tracers or non-tracer points \cite{2005MNRAS.361.1287M}. These correlations are either found to be consistent with zero or very small and then discarded. Or they are subtracted from the tangential shear to remove this potential systematic effect from the data.

The $\lambda$-statistics Eq.~\eqref{eq:dg_lambda_theory} introduced here are motivated in a similar way. They extend those previous works by quantifying the contribution of PSF-induced systematics for lensing by foreground density tracers. In particular, our formalism allows the propagation of PSF errors to derived physical quantities. Such quantities are, for example, the average halo mass of foreground galaxy samples, as illustrated in this paper. Other examples not studied here are the halo concentration, galaxy bias parameters, cluster mass calibration for cosmology, the intrinsic galaxy alignment amplitude, and void density profiles to name just a few.

We argue that residuals quantified by the $\lambda$-statistics are present in all weak-lensing measurements of density tracers. This is due to the spatially varying PSF within the field of view, a phenomenon that can lead to a correlation between the PSF and sky position.

Some of the PSF residuals can in principle be removed from the tangential shear from a density tracer by subtracting the shear around random points. However, this does now remove correlations between PSF and tracer number density.
This was noted in \cite{2005MNRAS.361.1287M}.
The $\lambda$-statistics quantify this correlation, and correspond to the term $\langle \delta n \, \gamma_\textrm{sys} \rangle$  in Eq.~(26) of \cite{2005MNRAS.361.1287M}. 

\subsection{Interpretation of the $\lambda$-statistic terms}
\label{sec:disc_lambda}

We now provide an intuitive interpretation of the three $\lambda$-statistic terms. The function $\lambda_1$ is qualitatively different from the other two, $\lambda_2$ and $\lambda_3$, which can be grouped together.

The first function, $\lambda_1$, quantifies the correlation between tangential PSF ellipticity and foreground sample position. The function $\lambda_1$ is proportional to the PSF ellipticity, not its residual, and can be of order $10\%$. For example, the requirement on the PSF ellipticity for the Euclid space mission is $0.15$ \citep{2011arXiv1110.3193L}; in UNIONS the focal-plane PSF ellipticity averaged over atmospheric fluctuations is of order $0.05$ \citep{UNIONS_Guinot_SP}.
Multiplied with a PSF leakage $\alpha$ at the percent-level \citep[e.g.]{2021A&A...645A.105G,10.1093/mnras/stab918} yields an additive PSF-induced bias, na\"lively of order $10^{-4}$ -- $10^{-3}$, which is indeed in the range of sought-after weak-lensing shear correlations.

For the $\lambda_1$-term to impact the measured tangential shear, the PSF ellipticity needs to display a spatial pattern that is correlated to the positions of the foreground sample.
An example is the commonly observed circular PSF pattern in the focal plane and the foreground sample being a cluster (sample) near the image centre. 
Note that the $\lambda_1$ term is independent of the quality of the PSF model.

The second function, $\lambda_2$, weighs the PSF ellipticity-foreground position correlation by the relative PSF size residual. This function shows some similarity to  $\rho_5 = \langle \varepsilon^\textrm{psf} (\varepsilon^{\textrm{psf}, \ast} \delta T^\textrm{psf} / T^\textrm{psf}) \rangle$ for cosmic shear, see App.~\ref{sec:rho_stats}.
The third function, $\lambda_3$, is the correlation of PSF ellipticity residuals with foreground positions, having a resemblance in $\rho_2 = \langle \varepsilon^\textrm{psf} \delta \varepsilon^{\textrm{psf}, *} \rangle$.
Both the second and third $\lambda$-statistics contribute to galaxy-galaxy lensing in the presence of PSF residuals that are spatially correlated to the positions of the foreground sample.
Such correlations can be induced where detection and selection of the density tracer populations depend on the PSF. Examples where PSF residuals can affect the tracer number density are:
\begin{itemize}
    \item seeing and depth variations;
    \item star-galaxy separation, cross-contamination of both samples;
    \item detector effects such as charge transfer inefficiency (CTI)
    \item survey strategy, e.g. scan direction, fiber placement
    \item photometry and photometric redshifts;
    \item local object density (crowded fields) and extinction
\end{itemize}
In all those cases, spurious correlations between PSF residuals and tracer positions might be introduced that will be captured by $\lambda_2$ and $\lambda_3$. Some of those effects such as varying seeing will only give rise to PSF - number density correlations if the lensing and density tracer catalogues originate from the same data. Others such as crowded fields are intrinsic. The contamination of the PSF sample by density tracers objects and vice versa can induce systematic biases studied here even if they are selected from different surveys if the target sample consists of small objects with size close to the PSF in either survey.

\subsection{Comparison to the $\rho$-statistics for cosmic shear}

Contrary to the $\rho$-statistics that were introduced to quantify PSF systematics for cosmic shear, the $\lambda$-statistics do not involve correlations between PSF uncertainties. 
Instead, they quantify the correlation of PSF residuals with respect to foreground positions.

The conditions for non-vanishing $\lambda$-statistics are stronger than for $\rho \ne 0$. In both cases, we need a non-perfect PSF model or PSF leakage that displays a spatial pattern. In addition, for the $\lambda$-statistics to be significant, the PSF residual pattern also needs to be correlated to the positions of the density tracer in question.

Related to this is the second difference between $\lambda$ and $\rho$: Cosmic shear involves second-order correlations of the statistical homogeneous and isotropic galaxy shear field. The $\rho$-statistics are evaluated not at galaxy but at star positions.
The star or PSF ellipticity field can be assumed to be statistically homogeneous and isotropic. This allows for the addition of the measured $\rho$ and $\xi_+$.

Galaxy-galaxy lensing however is the cross-correlation between two correlated fields, background shear and foreground density. The second-order correlation estimators are not invariant under translation
or rotation of only one of the fields. To capture PSF systematic correlations we need to use the actual galaxy positions.

\section{Summary}
\label{sec:summary}

This paper introduces PSF-induced systematics for weak-lensing cross-correlations with foreground density tracers. We developed a theoretical framework for quantifying PSF-induced systematics which contributes to both the multiplicative and the additive bias in weak-lensing tangential shear.
In particular, we introduce three scale-dependent ellipticity-position correlation functions dubbed ``$\lambda$-statistics'' to characterize PSF-induced additive biases. These correlation functions can be computed from the information about ellipticities and size residuals of the PSF, and the positions of the foreground sample.
In this framework, PSF-induced systematics propagate to physical parameters of the density tracer sample measured from weak lensing.

The PSF-induced multiplicative bias is a prefactor in the residual tangential shear. We use the UNIONS and DES weak-lensing catalogues to calculate PSF-induced multiplicative bias, which we find to be $-0.0042$ and $0.0003$, respectively. This is a subdominant contribution to the overall multiplicative bias of current weak-lensing surveys.

We quantify the impact of PSF-induced systematics on the halo masses of two cases of the foreground galaxy samples. The first case is a sample of central galaxies from the Yang group catalogue \citep{Yang2005, Yang2007}, which we split into subsamples by stellar mass and star-formation rate.
The PSF-induced bias acts mainly on small angular scales and low stellar masses. The largest resulting bias in the weak-lensing derived halo mass is $18\%$ for the subsample in the stellar mass range $9 \le \log M_\ast / M_\odot < 9.5$.

The second case is the type I AGN sample, used in \cite{2024arXiv240210740L} to estimate the black-hole-mass - halo-mass relation. We calculate the impact of PSF-induced systematics on the halo mass estimation.
Similar to the previous case, PSF-induced systematics is most important on small scales and absent on large scales. Without accounting for PSF-induced systematics, the halo mass is underestimated at low black-hole masses.

Our proposed framework can be used for quality-checking of weak-lensing - density cross-correlations. It is straightforward to extend the formalism to weak-lensing-like observables and estimators such as intrinsic alignments of galaxies.

\begin{acknowledgements}

This work was made possible by utilizing the CANDIDE cluster at the Institut d’Astrophysique de Paris, which was funded through grants from the PNCG, CNES, DIM-ACAV, and the Cosmic Dawn Center and maintained by S.~Rouberol.

We are honored and grateful for the opportunity of observing the Universe from Maunakea and Haleakala, which both have cultural, historical and natural significance in Hawai'i. This work is based on data obtained as part of the Canada-France Imaging Survey, a CFHT large program of the National Research Council of Canada and the French Centre National de la Recherche Scientifique. Based on observations obtained with MegaPrime/MegaCam, a joint project of CFHT and CEA Saclay, at the Canada-France-Hawaii Telescope (CFHT) which is operated by the National Research Council (NRC) of Canada, the Institut National des Science de l’Univers (INSU) of the Centre National de la Recherche Scientifique (CNRS) of France, and the University of Hawaii. This research used the facilities of the Canadian Astronomy Data Centre operated by the National Research Council of Canada with the support of the Canadian Space Agency. This research is based in part on data collected at Subaru Telescope, which is operated by the National Astronomical Observatory of Japan.
Pan-STARRS is a project of the Institute for Astronomy of the University of Hawai'i, and is supported by the NASA SSO Near Earth Observation Program under grants 80NSSC18K0971, NNX14AM74G, NNX12AR65G, NNX13AQ47G, NNX08AR22G, 80NSSC21K1572 and by the State of Hawai'i.

This work was supported in part by the Canadian Advanced Network for Astronomical Research (CANFAR) and Compute Canada facilities. 

This work was funded by the China Scholarship Council (CSC).

\end{acknowledgements}

\bibliographystyle{aa}
\bibliography{astro}

\begin{thebibliography}{54}
\expandafter\ifx\csname natexlab\endcsname\relax\def\natexlab#1{#1}\fi

\bibitem[{{Abazajian} {et~al.}(2009){Abazajian}, {Adelman-McCarthy},
  {Ag{\"u}eros}, {Allam}, {Allende Prieto}, {An}, {Anderson}, {Anderson}, \&
  {et al.}}]{Abazajian-09}
{Abazajian}, K.~N., {Adelman-McCarthy}, J.~K., {Ag{\"u}eros}, M.~A., {et~al.}
  2009, \apjs, 182, 543

\bibitem[{{Bhattacharya} {et~al.}(2013){Bhattacharya}, {Habib}, {Heitmann}, \&
  {Vikhlinin}}]{Bhattacharya2013ApJ}
{Bhattacharya}, S., {Habib}, S., {Heitmann}, K., \& {Vikhlinin}, A. 2013, \apj,
  766, 32

\bibitem[{{Bilicki} {et~al.}(2021){Bilicki}, {Dvornik}, {Hoekstra}, {Wright},
  {Chisari}, {Vakili}, {Asgari}, {Giblin}, {Heymans}, {Hildebrandt},
  {Holwerda}, {Hopkins}, {Johnston}, {Kannawadi}, {Kuijken}, {Nakoneczny},
  {Shan}, {Sonnenfeld}, \& {Valentijn}}]{Bilicki2021}
{Bilicki}, M., {Dvornik}, A., {Hoekstra}, H., {et~al.} 2021, \aap, 653, A82

\bibitem[{{Blanton} {et~al.}(2005){Blanton}, {Schlegel}, {Strauss},
  {Brinkmann}, {Finkbeiner}, {Fukugita}, {Gunn}, {Hogg}, \& {et
  al.}}]{Blanton-05a}
{Blanton}, M.~R., {Schlegel}, D.~J., {Strauss}, M.~A., {et~al.} 2005, \aj, 129,
  2562

\bibitem[{{Bluck} {et~al.}(2016){Bluck}, {Mendel}, {Ellison}, {Patton},
  {Simard}, {Henriques}, {Torrey}, {Teimoorinia}, \& {et al.}}]{Bluck2016}
{Bluck}, A.~F.~L., {Mendel}, J.~T., {Ellison}, S.~L., {et~al.} 2016, \mnras,
  462, 2559

\bibitem[{{Bower} {et~al.}(2017){Bower}, {Schaye}, {Frenk}, {Theuns},
  {Schaller}, {Crain}, \& {McAlpine}}]{Bower2017}
{Bower}, R.~G., {Schaye}, J., {Frenk}, C.~S., {et~al.} 2017, \mnras, 465, 32

\bibitem[{{Brinchmann} {et~al.}(2004){Brinchmann}, {Charlot}, {White},
  {Tremonti}, {Kauffmann}, {Heckman}, \& {Brinkmann}}]{Brinchmann2004}
{Brinchmann}, J., {Charlot}, S., {White}, S.~D.~M., {et~al.} 2004, \mnras, 351,
  1151

\bibitem[{{Diemer}(2018)}]{Diemer2018ApJS}
{Diemer}, B. 2018, \apjs, 239, 35

\bibitem[{{Euclid Collaboration} {et~al.}(2022){Euclid Collaboration},
  {Scaramella}, {Amiaux}, {Mellier}, {Burigana}, {Carvalho}, {Cuillandre}, {Da
  Silva}, {Derosa}, {Dinis}, {Maiorano}, {Maris}, {Tereno}, {Laureijs},
  {Boenke}, {Buenadicha}, {Dupac}, {Gaspar Venancio}, {G{\'o}mez-{\'A}lvarez},
  {Hoar}, {Lorenzo Alvarez}, {Racca}, {Saavedra-Criado}, {Schwartz}, {Vavrek},
  {Schirmer}, {Aussel}, {Azzollini}, {Cardone}, {Cropper}, {Ealet}, {Garilli},
  {Gillard}, {Granett}, {Guzzo}, {Hoekstra}, {Jahnke}, {Kitching}, {Maciaszek},
  {Meneghetti}, {Miller}, {Nakajima}, {Niemi}, {Pasian}, {Percival},
  {Pottinger}, {Sauvage}, {Scodeggio}, {Wachter}, {Zacchei}, {Aghanim},
  {Amara}, {Auphan}, {Auricchio}, {Awan}, {Balestra}, {Bender}, {Bodendorf},
  {Bonino}, {Branchini}, {Brau-Nogue}, {Brescia}, {Candini}, {Capobianco},
  {Carbone}, {Carlberg}, {Carretero}, {Casas}, {Castander}, {Castellano},
  {Cavuoti}, {Cimatti}, {Cledassou}, {Congedo}, {Conselice}, {Conversi},
  {Copin}, {Corcione}, {Costille}, {Courbin}, {Degaudenzi}, {Douspis},
  {Dubath}, {Duncan}, {Dusini}, {Farrens}, {Ferriol}, {Fosalba}, {Fourmanoit},
  {Frailis}, {Franceschi}, {Franzetti}, {Fumana}, {Gillis}, {Giocoli},
  {Grazian}, {Grupp}, {Haugan}, {Holmes}, {Hormuth}, {Hudelot}, {Kermiche},
  {Kiessling}, {Kilbinger}, {Kohley}, {Kubik}, {K{\"u}mmel}, {Kunz},
  {Kurki-Suonio}, {Lahav}, {Ligori}, {Lilje}, {Lloro}, {Mansutti}, {Marggraf},
  {Markovic}, {Marulli}, {Massey}, {Maurogordato}, {Melchior}, {Merlin},
  {Meylan}, {Mohr}, {Moresco}, {Morin}, {Moscardini}, {Munari}, {Nichol},
  {Padilla}, {Paltani}, {Peacock}, {Pedersen}, {Pettorino}, {Pires}, {Poncet},
  {Popa}, {Pozzetti}, {Raison}, {Rebolo}, {Rhodes}, {Rix}, {Roncarelli},
  {Rossetti}, {Saglia}, {Schneider}, {Schrabback}, {Secroun}, {Seidel},
  {Serrano}, {Sirignano}, {Sirri}, {Skottfelt}, {Stanco}, {Starck},
  {Tallada-Cresp{\'\i}}, {Tavagnacco}, {Taylor}, {Teplitz}, {Toledo-Moreo},
  {Torradeflot}, {Trifoglio}, {Valentijn}, {Valenziano}, {Verdoes Kleijn},
  {Wang}, {Welikala}, {Weller}, {Wetzstein}, {Zamorani}, {Zoubian}, {Andreon},
  {Baldi}, {Bardelli}, {Boucaud}, {Camera}, {Di Ferdinando}, {Fabbian},
  {Farinelli}, {Galeotta}, {Graci{\'a}-Carpio}, {Maino}, {Medinaceli}, {Mei},
  {Neissner}, {Polenta}, {Renzi}, {Romelli}, {Rosset}, {Sureau}, {Tenti},
  {Vassallo}, {Zucca}, {Baccigalupi}, {Balaguera-Antol{\'\i}nez}, {Battaglia},
  {Biviano}, {Borgani}, {Bozzo}, {Cabanac}, {Cappi}, {Casas}, {Castignani},
  {Colodro-Conde}, {Coupon}, {Courtois}, {Cuby}, {de la Torre}, {Desai},
  {Dole}, {Fabricius}, {Farina}, {Ferreira}, {Finelli}, {Flose-Reimberg},
  {Fotopoulou}, {Ganga}, {Gozaliasl}, {Hook}, {Keihanen}, {Kirkpatrick},
  {Liebing}, {Lindholm}, {Mainetti}, {Martinelli}, {Martinet}, {Maturi},
  {McCracken}, {Metcalf}, {Morgante}, {Nightingale}, {Nucita}, {Patrizii},
  {Potter}, {Riccio}, {S{\'a}nchez}, {Sapone}, {Schewtschenko}, {Schultheis},
  {Scottez}, {Teyssier}, {Tutusaus}, {Valiviita}, {Viel}, {Vriend}, \&
  {Whittaker}}]{2022A&A...662A.112E}
{Euclid Collaboration}, {Scaramella}, R., {Amiaux}, J., {et~al.} 2022, \aap,
  662, A112

\bibitem[{{Farrens} {et~al.}(2022){Farrens}, {Guinot}, {Kilbinger}, {Liaudat},
  {Baumont}, {Jimenez}, {Peel}, {Pujol}, {Schmitz}, {Starck}, \&
  {Vitorelli}}]{2022A&A...664A.141F}
{Farrens}, S., {Guinot}, A., {Kilbinger}, M., {et~al.} 2022, \aap, 664, A141

\bibitem[{{Foreman-Mackey} {et~al.}(2013){Foreman-Mackey}, {Hogg}, {Lang}, \&
  {Goodman}}]{Foreman-Mackey2013PASP}
{Foreman-Mackey}, D., {Hogg}, D.~W., {Lang}, D., \& {Goodman}, J. 2013, \pasp,
  125, 306

\bibitem[{Gatti {et~al.}(2021{\natexlab{a}})Gatti, Sheldon, Amon, Becker,
  Troxel, Choi, Doux, MacCrann, Navarro-Alsina, Harrison, Gruen, Bernstein,
  Jarvis, Secco, Ferté, Shin, McCullough, Rollins, Chen, Chang, Pandey,
  Tutusaus, Prat, Elvin-Poole, Sanchez, Plazas, Roodman, Zuntz, Abbott, Aguena,
  Allam, Annis, Avila, Bacon, Bertin, Bhargava, Brooks, Burke, Carnero Rosell,
  Carrasco Kind, Carretero, Castander, Conselice, Costanzi, Crocce, da Costa,
  Davis, De Vicente, Desai, Diehl, Dietrich, Doel, Drlica-Wagner, Eckert,
  Everett, Ferrero, Frieman, García-Bellido, Gerdes, Giannantonio, Gruendl,
  Gschwend, Gutierrez, Hartley, Hinton, Hollowood, Honscheid, Hoyle, Huff,
  Huterer, Jain, James, Jeltema, Krause, Kron, Kuropatkin, Lima, Maia,
  Marshall, Miquel, Morgan, Myles, Palmese, Paz-Chinchón, Rykoff, Samuroff,
  Sanchez, Scarpine, Schubnell, Serrano, Sevilla-Noarbe, Smith, Soares-Santos,
  Suchyta, Swanson, Tarle, Thomas, To, Tucker, Varga, Wechsler, Weller, Wester,
  \& Wilkinson}]{DES21_Gatti}
Gatti, M., Sheldon, E., Amon, A., {et~al.} 2021{\natexlab{a}}, Monthly Notices
  of the Royal Astronomical Society, 504, 4312

\bibitem[{Gatti {et~al.}(2021{\natexlab{b}})Gatti, Sheldon, Amon, Becker,
  Troxel, Choi, Doux, MacCrann, Navarro-Alsina, Harrison, Gruen, Bernstein,
  Jarvis, Secco, Ferté, Shin, McCullough, Rollins, Chen, Chang, Pandey,
  Tutusaus, Prat, Elvin-Poole, Sanchez, Plazas, Roodman, Zuntz, Abbott, Aguena,
  Allam, Annis, Avila, Bacon, Bertin, Bhargava, Brooks, Burke, Carnero Rosell,
  Carrasco Kind, Carretero, Castander, Conselice, Costanzi, Crocce, da Costa,
  Davis, De Vicente, Desai, Diehl, Dietrich, Doel, Drlica-Wagner, Eckert,
  Everett, Ferrero, Frieman, García-Bellido, Gerdes, Giannantonio, Gruendl,
  Gschwend, Gutierrez, Hartley, Hinton, Hollowood, Honscheid, Hoyle, Huff,
  Huterer, Jain, James, Jeltema, Krause, Kron, Kuropatkin, Lima, Maia,
  Marshall, Miquel, Morgan, Myles, Palmese, Paz-Chinchón, Rykoff, Samuroff,
  Sanchez, Scarpine, Schubnell, Serrano, Sevilla-Noarbe, Smith, Soares-Santos,
  Suchyta, Swanson, Tarle, Thomas, To, Tucker, Varga, Wechsler, Weller, Wester,
  \& Wilkinson}]{10.1093/mnras/stab918}
Gatti, M., Sheldon, E., Amon, A., {et~al.} 2021{\natexlab{b}}, Monthly Notices
  of the Royal Astronomical Society, 504, 4312

\bibitem[{{Giblin} {et~al.}(2021){Giblin}, {Heymans}, {Asgari}, {Hildebrandt},
  {Hoekstra}, {Joachimi}, {Kannawadi}, {Kuijken}, {Lin}, {Miller},
  {Tr{\"o}ster}, {van den Busch}, {Wright}, {Bilicki}, {Blake}, {de Jong},
  {Dvornik}, {Erben}, {Getman}, {Napolitano}, {Schneider}, {Shan}, \&
  {Valentijn}}]{2021A&A...645A.105G}
{Giblin}, B., {Heymans}, C., {Asgari}, M., {et~al.} 2021, \aap, 645, A105

\bibitem[{{Guinot} {et~al.}(2022){Guinot}, {Kilbinger}, {Farrens},
  {et~al.}}]{UNIONS_Guinot_SP}
{Guinot}, A., {Kilbinger}, M., {Farrens}, S., {et~al.} 2022, Submitted to
  \mnras

\bibitem[{{Guzik} \& {Seljak}(2002)}]{Guzik2002}
{Guzik}, J. \& {Seljak}, U. 2002, \mnras, 335, 311

\bibitem[{{Jarvis}(2015)}]{Jarvis2015}
{Jarvis}, M. 2015, {TreeCorr: Two-point correlation functions}, Astrophysics
  Source Code Library, record ascl:1508.007

\bibitem[{{Jarvis} {et~al.}(2021){Jarvis}, {Bernstein}, {Amon}, {Davis},
  {L{\'e}get}, {Bechtol}, {Harrison}, {Gatti}, {Roodman}, {Chang}, {Chen},
  {Choi}, {Desai}, {Drlica-Wagner}, {Gruen}, {Gruendl}, {Hernandez},
  {MacCrann}, {Meyers}, {Navarro-Alsina}, {Pandey}, {Plazas}, {Secco},
  {Sheldon}, {Troxel}, {Vorperian}, {Wei}, {Zuntz}, {Abbott}, {Aguena},
  {Allam}, {Avila}, {Bhargava}, {Bridle}, {Brooks}, {Carnero Rosell}, {Carrasco
  Kind}, {Carretero}, {Costanzi}, {da Costa}, {De Vicente}, {Diehl}, {Doel},
  {Everett}, {Flaugher}, {Fosalba}, {Frieman}, {Garc{\'\i}a-Bellido},
  {Gaztanaga}, {Gerdes}, {Gutierrez}, {Hinton}, {Hollowood}, {Honscheid},
  {James}, {Kent}, {Kuehn}, {Kuropatkin}, {Lahav}, {Maia}, {March}, {Marshall},
  {Melchior}, {Menanteau}, {Miquel}, {Ogando}, {Paz-Chinch{\'o}n}, {Rykoff},
  {Sanchez}, {Scarpine}, {Schubnell}, {Serrano}, {Sevilla-Noarbe}, {Smith},
  {Suchyta}, {Swanson}, {Tarle}, {Varga}, {Walker}, {Wester}, {Wilkinson},
  {Wilkinson}, \& {DES Collaboration}}]{Jarvis2021}
{Jarvis}, M., {Bernstein}, G.~M., {Amon}, A., {et~al.} 2021, \mnras, 501, 1282

\bibitem[{{Jarvis} {et~al.}(2016){Jarvis}, {Sheldon}, {Zuntz}, {Kacprzak},
  {Bridle}, {Amara}, {Armstrong}, {Becker}, {Bernstein}, {Bonnett}, {Chang},
  {Das}, {Dietrich}, {Drlica-Wagner}, {Eifler}, {Gangkofner}, {Gruen},
  {Hirsch}, {Huff}, {Jain}, {Kent}, {Kirk}, {MacCrann}, {Melchior}, {Plazas},
  {Refregier}, {Rowe}, {Rykoff}, {Samuroff}, {S{\'a}nchez}, {Suchyta},
  {Troxel}, {Vikram}, {Abbott}, {Abdalla}, {Allam}, {Annis}, {Benoit-L{\'e}vy},
  {Bertin}, {Brooks}, {Buckley-Geer}, {Burke}, {Capozzi}, {Carnero Rosell},
  {Carrasco Kind}, {Carretero}, {Castander}, {Clampitt}, {Crocce}, {Cunha},
  {D'Andrea}, {da Costa}, {DePoy}, {Desai}, {Diehl}, {Doel}, {Fausti Neto},
  {Flaugher}, {Fosalba}, {Frieman}, {Gaztanaga}, {Gerdes}, {Gruendl},
  {Gutierrez}, {Honscheid}, {James}, {Kuehn}, {Kuropatkin}, {Lahav}, {Li},
  {Lima}, {March}, {Martini}, {Miquel}, {Mohr}, {Neilsen}, {Nord}, {Ogando},
  {Reil}, {Romer}, {Roodman}, {Sako}, {Sanchez}, {Scarpine}, {Schubnell},
  {Sevilla-Noarbe}, {Smith}, {Soares-Santos}, {Sobreira}, {Swanson}, {Tarle},
  {Thaler}, {Thomas}, {Walker}, \& {Wechsler}}]{Jarvis2016}
{Jarvis}, M., {Sheldon}, E., {Zuntz}, J., {et~al.} 2016, \mnras, 460, 2245

\bibitem[{{Kilbinger}(2015)}]{K15}
{Kilbinger}, M. 2015, Reports on Progress in Physics, 78, 086901

\bibitem[{{Kravtsov} {et~al.}(2018){Kravtsov}, {Vikhlinin}, \&
  {Meshcheryakov}}]{Kravtsov2018}
{Kravtsov}, A.~V., {Vikhlinin}, A.~A., \& {Meshcheryakov}, A.~V. 2018,
  Astronomy Letters, 44, 8

\bibitem[{{Laureijs} {et~al.}(2011){Laureijs}, {Amiaux}, {Arduini},
  {Augu{\`e}res}, {Brinchmann}, {Cole}, {Cropper}, {Dabin}, {Duvet}, {Ealet},
  {et~al.}}]{2011arXiv1110.3193L}
{Laureijs}, R., {Amiaux}, J., {Arduini}, S., {et~al.} 2011, arXiv:1110.3193
  [\eprint[arXiv]{1110.3193}]

\bibitem[{{Leauthaud} {et~al.}(2012){Leauthaud}, {Tinker}, {Bundy}, {Behroozi},
  {Massey}, {Rhodes}, {George}, {Kneib}, {Benson}, {Wechsler}, {Busha},
  {Capak}, {Cort{\^e}s}, {Ilbert}, {Koekemoer}, {Le F{\`e}vre}, {Lilly},
  {McCracken}, {Salvato}, {Schrabback}, {Scoville}, {Smith}, \&
  {Taylor}}]{Leauthaud2012}
{Leauthaud}, A., {Tinker}, J., {Bundy}, K., {et~al.} 2012, \apj, 744, 159

\bibitem[{{Li} {et~al.}(2024){Li}, {Kilbinger}, {Luo}, {Wang}, {Wang},
  {Wittje}, {Hildebrandt}, {van Waerbeke}, {Hudson}, {Farrens}, {Liaudat},
  {Liu}, {Zhang}, {Wang}, {Russier}, {Guinot}, {Baumont}, {Hervas Peters}, {de
  Boer}, \& {Wang}}]{2024arXiv240210740L}
{Li}, Q., {Kilbinger}, M., {Luo}, W., {et~al.} 2024, submitted to ApJL,
  arXiv:2402.10740

\bibitem[{{Liaudat} {et~al.}(2021){Liaudat}, {Bonnin}, {Starck}, {Schmitz},
  {Guinot}, {Kilbinger}, \& {Gwyn}}]{MCCD21}
{Liaudat}, T., {Bonnin}, J., {Starck}, J.-L., {et~al.} 2021, \aap, 646, A27

\bibitem[{{Liaudat} {et~al.}(2023){Liaudat}, {Starck}, \&
  {Kilbinger}}]{2023FrASS..1058213L}
{Liaudat}, T.~I., {Starck}, J.-L., \& {Kilbinger}, M. 2023, Frontiers in
  Astronomy and Space Sciences, 10, 1158213

\bibitem[{{Liu} {et~al.}(2019){Liu}, {Liu}, {Dong}, {Zhou}, {Wang}, {Lu}, \&
  {Yuan}}]{Liu2019}
{Liu}, H.-Y., {Liu}, W.-J., {Dong}, X.-B., {et~al.} 2019, \apjs, 243, 21

\bibitem[{{Luo} {et~al.}(2018){Luo}, {Yang}, {Lu}, {Shi}, {Zhang}, {Mo}, {Shu},
  {Fu}, {Radovich}, {Zhang}, {Li}, {Sunayama}, \& {Wang}}]{Luo2018ApJ}
{Luo}, W., {Yang}, X., {Lu}, T., {et~al.} 2018, \apj, 862, 4

\bibitem[{{Lyke} {et~al.}(2020){Lyke}, {Higley}, {McLane}, {Schurhammer},
  {Myers}, {Ross}, {Dawson}, {Chabanier}, {Martini}, {Busca}, {Mas des
  Bourboux}, {Salvato}, {Streblyanska}, {Zarrouk}, {Burtin}, {Anderson},
  {Bautista}, {Bizyaev}, {Brandt}, {Brinkmann}, {Brownstein}, {Comparat},
  {Green}, {de la Macorra}, {Mu{\~n}oz Guti{\'e}rrez}, {Hou}, {Newman},
  {Palanque-Delabrouille}, {P{\^a}ris}, {Percival}, {Petitjean}, {Rich},
  {Rossi}, {Schneider}, {Smith}, {Vivek}, \& {Weaver}}]{Lyke2020}
{Lyke}, B.~W., {Higley}, A.~N., {McLane}, J.~N., {et~al.} 2020, \apjs, 250, 8

\bibitem[{{MacCrann} {et~al.}(2022){MacCrann}, {Becker}, {McCullough}, {Amon},
  {Gruen}, {Jarvis}, {Choi}, {Troxel}, {Sheldon}, {Yanny}, {Herner},
  {Dodelson}, {Zuntz}, {Eckert}, {Rollins}, {Varga}, {Bernstein}, {Gruendl},
  {Harrison}, {Hartley}, {Sevilla-Noarbe}, {Pieres}, {Bridle}, {Myles},
  {Alarcon}, {Everett}, {S{\'a}nchez}, {Huff}, {Tarsitano}, {Gatti}, {Secco},
  {Abbott}, {Aguena}, {Allam}, {Annis}, {Bacon}, {Bertin}, {Brooks}, {Burke},
  {Carnero Rosell}, {Carrasco Kind}, {Carretero}, {Costanzi}, {Crocce},
  {Pereira}, {De Vicente}, {Desai}, {Diehl}, {Dietrich}, {Doel}, {Eifler},
  {Ferrero}, {Fert{\'e}}, {Flaugher}, {Fosalba}, {Frieman},
  {Garc{\'\i}a-Bellido}, {Gaztanaga}, {Gerdes}, {Giannantonio}, {Gschwend},
  {Gutierrez}, {Hinton}, {Hollowood}, {Honscheid}, {James}, {Lahav}, {Lima},
  {Maia}, {March}, {Marshall}, {Martini}, {Melchior}, {Menanteau}, {Miquel},
  {Mohr}, {Morgan}, {Muir}, {Ogando}, {Palmese}, {Paz-Chinch{\'o}n}, {Plazas},
  {Rodriguez-Monroy}, {Roodman}, {Samuroff}, {Sanchez}, {Scarpine}, {Serrano},
  {Smith}, {Soares-Santos}, {Suchyta}, {Swanson}, {Tarle}, {Thomas}, {To},
  {Wilkinson}, {Wilkinson}, \& {DES Collaboration}}]{2022MNRAS.509.3371M}
{MacCrann}, N., {Becker}, M.~R., {McCullough}, J., {et~al.} 2022, \mnras, 509,
  3371

\bibitem[{{Mandelbaum}(2018)}]{2017arXiv171003235M}
{Mandelbaum}, R. 2018, \araa, 56, 393

\bibitem[{{Mandelbaum} {et~al.}(2005){Mandelbaum}, {Hirata}, {Seljak}, {Guzik},
  {Padmanabhan}, {Blake}, {Blanton}, {Lupton}, \&
  {Brinkmann}}]{2005MNRAS.361.1287M}
{Mandelbaum}, R., {Hirata}, C.~M., {Seljak}, U., {et~al.} 2005, \mnras, 361,
  1287

\bibitem[{{Mandelbaum} {et~al.}(2006){Mandelbaum}, {Seljak}, {Kauffmann},
  {Hirata}, \& {Brinkmann}}]{Mandelbaum2006}
{Mandelbaum}, R., {Seljak}, U., {Kauffmann}, G., {Hirata}, C.~M., \&
  {Brinkmann}, J. 2006, \mnras, 368, 715

\bibitem[{{Mandelbaum} {et~al.}(2016){Mandelbaum}, {Wang}, {Zu}, {White},
  {Henriques}, \& {More}}]{Mandelbaum2016}
{Mandelbaum}, R., {Wang}, W., {Zu}, Y., {et~al.} 2016, \mnras, 457, 3200

\bibitem[{{Massey} {et~al.}(2013){Massey}, {Hoekstra}, {Kitching}, {Rhodes},
  {Cropper}, {Amiaux}, {Harvey}, {Mellier}, {Meneghetti}, {Miller},
  {Paulin-Henriksson}, {Pires}, {Scaramella}, \& {Schrabback}}]{Massey2013}
{Massey}, R., {Hoekstra}, H., {Kitching}, T., {et~al.} 2013, \mnras, 429, 661

\bibitem[{{Moster} {et~al.}(2010){Moster}, {Somerville}, {Maulbetsch}, {van den
  Bosch}, {Macci{\`o}}, {Naab}, \& {Oser}}]{Moster2010}
{Moster}, B.~P., {Somerville}, R.~S., {Maulbetsch}, C., {et~al.} 2010, \apj,
  710, 903

\bibitem[{{Navarro} {et~al.}(1997){Navarro}, {Frenk}, \& {White}}]{Navarro1997}
{Navarro}, J.~F., {Frenk}, C.~S., \& {White}, S. D.~M. 1997, \apj, 490, 493

\bibitem[{{Paulin-Henriksson} {et~al.}(2008){Paulin-Henriksson}, {Amara},
  {Voigt}, {Refregier}, \& {Bridle}}]{Henriksson2008}
{Paulin-Henriksson}, S., {Amara}, A., {Voigt}, L., {Refregier}, A., \&
  {Bridle}, S.~L. 2008, \aap, 484, 67

\bibitem[{{Planck Collaboration} {et~al.}(2016){Planck Collaboration}, {Ade},
  {Aghanim}, {Arnaud}, {Ashdown}, {Aumont}, {Baccigalupi}, {Banday},
  {Barreiro}, {Bartlett}, {Bartolo}, {Battaner}, {Battye}, {Benabed},
  {Beno{\^\i}t}, {Benoit-L{\'e}vy}, {Bernard}, {Bersanelli}, {Bielewicz},
  {Bock}, {Bonaldi}, {Bonavera}, {Bond}, {Borrill}, {Bouchet}, {Boulanger},
  {Bucher}, {Burigana}, {Butler}, {Calabrese}, {Cardoso}, {Catalano},
  {Challinor}, {Chamballu}, {Chary}, {Chiang}, {Chluba}, {Christensen},
  {Church}, {Clements}, {Colombi}, {Colombo}, {Combet}, {Coulais}, {Crill},
  {Curto}, {Cuttaia}, {Danese}, {Davies}, {Davis}, {de Bernardis}, {de Rosa},
  {de Zotti}, {Delabrouille}, {D{\'e}sert}, {Di Valentino}, {Dickinson},
  {Diego}, {Dolag}, {Dole}, {Donzelli}, {Dor{\'e}}, {Douspis}, {Ducout},
  {Dunkley}, {Dupac}, {Efstathiou}, {Elsner}, {En{\ss}lin}, {Eriksen},
  {Farhang}, {Fergusson}, {Finelli}, {Forni}, {Frailis}, {Fraisse},
  {Franceschi}, {Frejsel}, {Galeotta}, {Galli}, {Ganga}, {Gauthier}, {Gerbino},
  {Ghosh}, {Giard}, {Giraud-H{\'e}raud}, {Giusarma}, {Gjerl{\o}w},
  {Gonz{\'a}lez-Nuevo}, {G{\'o}rski}, {Gratton}, {Gregorio}, {Gruppuso},
  {Gudmundsson}, {Hamann}, {Hansen}, {Hanson}, {Harrison}, {Helou},
  {Henrot-Versill{\'e}}, {Hern{\'a}ndez-Monteagudo}, {Herranz}, {Hildebrandt},
  {Hivon}, {Hobson}, {Holmes}, {Hornstrup}, {Hovest}, {Huang}, {Huffenberger},
  {Hurier}, {Jaffe}, {Jaffe}, {Jones}, {Juvela}, {Keih{\"a}nen}, {Keskitalo},
  {Kisner}, {Kneissl}, {Knoche}, {Knox}, {Kunz}, {Kurki-Suonio}, {Lagache},
  {L{\"a}hteenm{\"a}ki}, {Lamarre}, {Lasenby}, {Lattanzi}, {Lawrence}, {Leahy},
  {Leonardi}, {Lesgourgues}, {Levrier}, {Lewis}, {Liguori}, {Lilje},
  {Linden-V{\o}rnle}, {L{\'o}pez-Caniego}, {Lubin}, {Mac{\'\i}as-P{\'e}rez},
  {Maggio}, {Maino}, {Mandolesi}, {Mangilli}, {Marchini}, {Maris}, {Martin},
  {Martinelli}, {Mart{\'\i}nez-Gonz{\'a}lez}, {Masi}, {Matarrese}, {McGehee},
  {Meinhold}, {Melchiorri}, {Melin}, {Mendes}, {Mennella}, {Migliaccio},
  {Millea}, {Mitra}, {Miville-Desch{\^e}nes}, {Moneti}, {Montier}, {Morgante},
  {Mortlock}, {Moss}, {Munshi}, {Murphy}, {Naselsky}, {Nati}, {Natoli},
  {Netterfield}, {N{\o}rgaard-Nielsen}, {Noviello}, {Novikov}, {Novikov},
  {Oxborrow}, {Paci}, {Pagano}, {Pajot}, {Paladini}, {Paoletti}, {Partridge},
  {Pasian}, {Patanchon}, {Pearson}, {Perdereau}, {Perotto}, {Perrotta},
  {Pettorino}, {Piacentini}, {Piat}, {Pierpaoli}, {Pietrobon}, {Plaszczynski},
  {Pointecouteau}, {Polenta}, {Popa}, {Pratt}, {Pr{\'e}zeau}, {Prunet},
  {Puget}, {Rachen}, {Reach}, {Rebolo}, {Reinecke}, {Remazeilles}, {Renault},
  {Renzi}, {Ristorcelli}, {Rocha}, {Rosset}, {Rossetti}, {Roudier},
  {Rouill{\'e} d'Orfeuil}, {Rowan-Robinson}, {Rubi{\~n}o-Mart{\'\i}n},
  {Rusholme}, {Said}, {Salvatelli}, {Salvati}, {Sandri}, {Santos},
  {Savelainen}, {Savini}, {Scott}, {Seiffert}, {Serra}, {Shellard}, {Spencer},
  {Spinelli}, {Stolyarov}, {Stompor}, {Sudiwala}, {Sunyaev}, {Sutton},
  {Suur-Uski}, {Sygnet}, {Tauber}, {Terenzi}, {Toffolatti}, {Tomasi},
  {Tristram}, {Trombetti}, {Tucci}, {Tuovinen}, {T{\"u}rler}, {Umana},
  {Valenziano}, {Valiviita}, {Van Tent}, {Vielva}, {Villa}, {Wade}, {Wandelt},
  {Wehus}, {White}, {White}, {Wilkinson}, {Yvon}, {Zacchei}, \&
  {Zonca}}]{Planck2016}
{Planck Collaboration}, {Ade}, P.~A.~R., {Aghanim}, N., {et~al.} 2016, \aap,
  594, A13

\bibitem[{{Posti} {et~al.}(2019){Posti}, {Fraternali}, \&
  {Marasco}}]{Posti2019}
{Posti}, L., {Fraternali}, F., \& {Marasco}, A. 2019, \aap, 626, A56

\bibitem[{{Rowe}(2010)}]{2010MNRAS.404..350R}
{Rowe}, B. 2010, \mnras, 404, 350

\bibitem[{{Shankar} {et~al.}(2020){Shankar}, {Allevato}, {Bernardi}, {Marsden},
  {Lapi}, {Menci}, {Grylls}, {Krumpe}, {Zanisi}, {Ricci}, {La Franca}, {Baldi},
  {Moreno}, \& {Sheth}}]{Shankar2020}
{Shankar}, F., {Allevato}, V., {Bernardi}, M., {et~al.} 2020, Nature Astronomy,
  4, 282

\bibitem[{{Tinker} {et~al.}(2010){Tinker}, {Robertson}, {Kravtsov}, {Klypin},
  {Warren}, {Yepes}, \& {Gottl{\"o}ber}}]{Tinker2010}
{Tinker}, J.~L., {Robertson}, B.~E., {Kravtsov}, A.~V., {et~al.} 2010, \apj,
  724, 878

\bibitem[{{van Uitert} \& {Schneider}(2016)}]{Uitert2016}
{van Uitert}, E. \& {Schneider}, P. 2016, \aap, 595, A93

\bibitem[{{Velander} {et~al.}(2014){Velander}, {van Uitert}, {Hoekstra},
  {Coupon}, {Erben}, {Heymans}, {Hildebrandt}, {Kitching}, {Mellier}, {Miller},
  {Van Waerbeke}, {Bonnett}, {Fu}, {Giodini}, {Hudson}, {Kuijken}, {Rowe},
  {Schrabback}, \& {Semboloni}}]{Velander2014MNRAS.437.2111V}
{Velander}, M., {van Uitert}, E., {Hoekstra}, H., {et~al.} 2014, \mnras, 437,
  2111

\bibitem[{{Viola} {et~al.}(2015){Viola}, {Cacciato}, {Brouwer}, {Kuijken},
  {Hoekstra}, {Norberg}, {Robotham}, {van Uitert}, {Alpaslan}, {Baldry},
  {Choi}, {de Jong}, {Driver}, {Erben}, {Grado}, {Graham}, {Heymans},
  {Hildebrandt}, {Hopkins}, {Irisarri}, {Joachimi}, {Loveday}, {Miller},
  {Nakajima}, {Schneider}, {Sif{\'o}n}, \& {Verdoes
  Kleijn}}]{Viola2015MNRAS.452.3529V}
{Viola}, M., {Cacciato}, M., {Brouwer}, M., {et~al.} 2015, \mnras, 452, 3529

\bibitem[{{Wu} \& {Shen}(2022)}]{Wu2022}
{Wu}, Q. \& {Shen}, Y. 2022, \apjs, 263, 42

\bibitem[{{Yang} {et~al.}(2009){Yang}, {Mo}, \& {van den Bosch}}]{Yang2009}
{Yang}, X., {Mo}, H.~J., \& {van den Bosch}, F.~C. 2009, \apj, 695, 900

\bibitem[{{Yang} {et~al.}(2005){Yang}, {Mo}, {van den Bosch}, \&
  {Jing}}]{Yang2005}
{Yang}, X., {Mo}, H.~J., {van den Bosch}, F.~C., \& {Jing}, Y.~P. 2005, \mnras,
  356, 1293

\bibitem[{{Yang} {et~al.}(2006){Yang}, {Mo}, {van den Bosch}, {Jing},
  {Weinmann}, \& {Meneghetti}}]{Yang2006}
{Yang}, X., {Mo}, H.~J., {van den Bosch}, F.~C., {et~al.} 2006, \mnras, 373,
  1159

\bibitem[{{Yang} {et~al.}(2007){Yang}, {Mo}, {van den Bosch}, {Pasquali}, {Li},
  \& {Barden}}]{Yang2007}
{Yang}, X., {Mo}, H.~J., {van den Bosch}, F.~C., {et~al.} 2007, \apj, 671, 153

\bibitem[{{Zhang} {et~al.}(2024){Zhang}, {Wang}, {Luo}, {Mo}, {Zhang}, {Yang},
  {Li}, \& {Li}}]{zhang2024}
{Zhang}, Z., {Wang}, H., {Luo}, W., {et~al.} 2024, \apj, 960, 71

\bibitem[{{Zhang} {et~al.}(2021){Zhang}, {Wang}, {Luo}, {Mo}, {Liang}, {Li},
  {Yang}, {Wang}, {Zhang}, {Hong}, {Wang}, {Wang}, {Li}, \& {Shi}}]{Zhang2021}
{Zhang}, Z., {Wang}, H., {Luo}, W., {et~al.} 2021, \aap, 650, A155

\bibitem[{{Zhang} {et~al.}(2022){Zhang}, {Wang}, {Luo}, {Zhang}, {Mo}, {Jing},
  {Yang}, \& {Li}}]{Zhang2022}
{Zhang}, Z., {Wang}, H., {Luo}, W., {et~al.} 2022, \aap, 663, A85

\end{thebibliography}

\begin{appendix}

\section{$\rho$-statistics for cosmic shear}
\label{sec:rho_stats}

Two correlation functions between PSF ellipticity and its residuals were introduced in \cite{2010MNRAS.404..350R}. The purpose of those diagnostics was to distinguish between different PSF models in a quantitative way  In particular, the second correlation function, $D_2$, corresponding to $\rho_2$ in the \citet{Jarvis2016} notation, is an indication of over-fitting. In that case the PSF model fits part of the noise, which creates a correlation between the observed PSF ellipticity and the model (residual).

These correlations were generalised in \cite{Jarvis2016}. This work rederived the two original \cite{2010MNRAS.404..350R} diagnostics and three additional functions by considering the two-point correlator of Eq.~\eqref{eq:epsobs}, which is an estimator of the shear two-point correlation function,
\begin{equation}
   \hat \xi_+(\theta) = \left\langle \varepsilon^\textrm{obs} \varepsilon^{\textrm{obs}, \ast} \right\rangle(\theta) ,
\end{equation}
into which Eq.~\eqref{eq:delta_eps} is inserted. The $\rho$-statistics are defined as second-order correlation functions
\begin{align}
    \rho_1(\theta) & = \left\langle
        \delta \varepsilon^\PSF \, \delta \varepsilon^{\PSF, \ast}
        \right\rangle(\theta) ;
        \nonumber \\
    \rho_2(\theta) & = \left\langle
        \varepsilon^\PSF \, \delta \varepsilon^{\PSF, \ast}
        \right\rangle(\theta) ;
        \nonumber \\
    \rho_3(\theta) & = \left\langle
        \left( \varepsilon^\PSF \frac{\delta T^\PSF}{T^\PSF} \right) 
        \left( \varepsilon^{\PSF, \ast} \frac{\delta T^\PSF}{T^\PSF} \right)
        \right\rangle(\theta) ;
        \nonumber \\
    \rho_4(\theta) & = \left\langle
        \delta \varepsilon^\PSF
        \left( \varepsilon^{\PSF, \ast} \frac{\delta T^\PSF}{T^\PSF}\right)
        \right\rangle(\theta) ;
        \nonumber \\
    \rho_5(\theta) & =  \left\langle
        \varepsilon^\PSF
        \left( \varepsilon^{\PSF, \ast} \frac{\delta T^\PSF}{T^\PSF}\right)
        \right\rangle(\theta) .
    \label{eq:rho5}
\end{align}
With some prefactors that depend on the ratio between PSF and galaxy size and the leakage parameter $\alpha$, the five $\rho$-statistic functions are written as additive terms to the cosmological shear two-point correlation function $\xi_+$, estimated by the correlation of observed galaxy ellipticities.


The choice to leave the size ratio and leakage parameters out of the $\rho$-statistics is conveniently done such that the quantities to correlate only depend on quantities that are available at star positions. 
In \cite{2010MNRAS.404..350R} the size ratio and leakage parameter are computed independently from the $\rho$-statistics.

The ensemble average of the $\rho$-statistics is thus estimated using star position, whereas the two-point correlation function $\xi_+$ is estimated by averaging over galaxy positions. Both can be added if both stars and galaxies randomly sample the underlying PSF and cosmic-shear fields, respectively.


The case of galaxy-galaxy lensing is more complex since the $\lambda$-statistics involve star - galaxy cross-correlations.

\end{appendix}

\end{document}